\documentclass[useAMS,usenatbib]{mn2e}
\usepackage{times}
\usepackage{graphicx}
\usepackage{epstopdf}
\usepackage{amssymb,amsmath}
\newcounter{subfigure}

\title[Pulsating sdB in the second half of the survey phase.]{First {\it Kepler} results on compact pulsators \\VII. Pulsating subdwarf B stars detected in the second half of the survey phase}
\author[A. S. Baran et al.]
{A. S. Baran $^{1,2}$\thanks{E-mail:asb@iastate.edu},
S. D. Kawaler $^{1}$, 
M. D. Reed$^{3}$,
A. C. Quint$^{3}$,
S. J. O'Toole$^{4}$,
\newauthor
R. H. \O stensen$^{5}$,
J. H. Telting$^{6}$
R. Silvotti$^{7}$,
S. Charpinet$^{8}$,
\newauthor
J. Christensen-Dalsgaard$^{9}$,
M. Still$^{10}$,
J.R. Hall$^{11}$ and
K. Uddin$^{11}$ 
\\
$^{1}$ Iowa State University, Department of Physics and Astronomy, 12 Physics Hall, Ames, IA 50011, USA\\
$^{2}$ Mt. Suhora Observatory, Cracow Pedagogical University, ul. Podchorazych 2, 30-084 Krakow, Poland\\
$^{3}$ Department of Physics, Astronomy, and Materials Science, Missouri State University, Springfield, MO 65897, USA\\
$^{4}$ Anglo-Australian Observatory, PO Box 296, Epping NSW 1710, Australia\\
$^{5}$ Instituut voor Sterrenkunde, K.U.~Leuven, Celestijnenlaan 200D, 3001 Leuven, Belgium\\
$^{6}$ Nordic Optical Telescope, 38700 Santa Cruz de La Palma, Spain\\
$^{7}$INAF-Osservatorio Astronomico di Torino, Strada dell'Osservatorio 20, 10025 Pino Torinese, Italy\\
$^{8}$Laboratoire d'Astrophysique de Toulouse-Tarbes, Universit\'e de Toulouse, CNRS, 14 Av. E. Belin, 31400 Toulouse, France\\
$^{9}$ Department of Physics and Astronomy, Aarhus University, 8000 Aarhus C, Denmark\\
$^{10}$ Bay Area Environmental Research Inst./NASA Ames Research Center, Moffett Field, CA 94035, USA\\
$^{11}$ Orbital Sciences Corporation/NASA Ames Research Center, Moffett Field, CA 94035, USA
}
\begin{document}

\date{}

\pagerange{\pageref{firstpage}--\pageref{lastpage}} \pubyear{2010}

\maketitle

\label{firstpage}

\begin{abstract}
We present five new pulsating subdwarf B (sdB) stars discovered by the {\em Kepler} spacecraft during the asteroseismology survey phase. We perform time-series analysis on the nearly continuous month-long {\em Kepler} datasets of these 5 objects; these datasets provide nearly alias--free time--series photometry at unprecedented precision. Following an iterative prewhitening process we derive the pulsational frequency spectra of these stars, separating out artefacts of known instrumental 
origin. We find that these new pulsating sdB stars are multiperiodic long-period pulsators of the V1093\,Her type, with the number of periodicities ranging from 8 (KIC\,8302197) to 53 (KIC\,11558725). The frequencies and amplitudes are typical of $g$-mode pulsators of this type. We do not find any evidence for binarity in the five stars from their observed pulsation frequencies. As these are $g-$mode pulsators, we briefly looked for period spacings  for mode identification, and found average spacings about 260\,s and 145\,s. This may indicate $l$\,=\,1 and 2 patterns. Some modes may show evidence of rotational splitting.
These discoveries complete the list of compact pulsators found in the survey phase. Of the 13 compact pulsators, only one star was identified as a short-period ($p$-mode) V361\,Hya pulsator, while all other new pulsators turned out to be V1093\,Her class objects. Among the latter objects, two of them seemed to be pure V1093\,Her while the others show additional low amplitude peaks in the $p-$mode frequency range, suggesting their hybrid nature. Authenticity of these peaks will be tested with longer runs currently under analysis.
\end{abstract}


\section{Introduction}
\label{introduction}
Hot subdwarf B (hereafter: sdB) stars are horizontal branch stars that consist of a helium rich core surrounded by a thin hydrogen envelope. The helium core sustains nuclear burning of helium into carbon and oxygen, though in some cases the core may have exhausted helium at the center. Their progenitors are low-mass main sequence stars that are likely to have undergone the core helium flash ($M \lesssim 2 M_{\odot}$). Once hydrogen in the core is exhausted, their progenitors evolve towards the tip of the red giant branch, where they lose all but a tiny fraction of hydrogen during (or prior to) the helium flash.  Those stars that retain less than about 0.001$M_{\odot}$ of hydrogen lie on the high $T_{\rm eff}$ end of the horizontal branch. This small fraction of hydrogen has an influence on the future evolution of sdB stars. As hydrogen shell burning cannot be sustained by the thin hydrogen envelope, the objects will move directly to the white dwarfs cooling track, instead of going through the asymptotic giant branch phase. Several mechanisms for producing this small residual hydrogen envelope that involve single-star or binary evolution have been proposed;  e.g. \cite{dcruz96} and \cite{han02,han03}.
The hot subdwarfs have representative effective temperatures and surface gravities near 30\,000~K and $\log g \sim$\,5.5, respectively. For a general review of their observed properties, see \cite{heber09}.

\begin{table*}
 \centering
 \begin{minipage}{83mm}
  \caption{Log of observations.}
  \label{phot_log}
  \begin{tabular}{@{}ccccccc@{}}
  \hline
  KIC           &  start            & end              &  length & freq. resolution & points & roll \\
 \hline
       &  \multicolumn{2}{c}{BJD-2455000} & days  & $\mu$Hz & & \\
\hline
 7668647  &  93.21388  & 123.55591  &  30.3  & 0.33  & 43050 & 3.1 \\
10001893 & 124.41490 & 154.45049 &  30.0   & 0.33 & 43932 & 3.2 \\
 8302197  & 156.51503 & 182.50487 &  26.0   & 0.38 & 37997 & 3.3 \\
11558725 & 156.51526 & 182.50527 &  26.0   & 0.38 & 38020 & 3.3 \\
10553698 & 185.36665 & 216.50590 &  31.1   & 0.32 & 45467 & 4.1 \\
\hline
\end{tabular}
\end{minipage}
\end{table*}

\citet{kilkenny97} discovered the first hot subdwarf that showed stellar oscillations (hereafter: sdBV). The discovery provided the first opportunity to use asteroseismology to look inside these enigmatic stars, with many more found with similar periods in the following years.  Now, around 50 objects of this class are known \citep{ostensen10a}. At the same time, independent theoretical work suggested that these stars could undergo nonradial pulsations \citep{charp96}, providing a ``jump-start'' to exploiting their asteroseismic potential.  The first pulsators showed relatively short--period pulsations (the short-period sdBV stars, or V361\,Hya stars, frequencies higher than about 2000$\mu$Hz); these are attributed to pressure modes ($p$-modes) and are driven in the outer part of the star.  As discovered later, sdB stars can also show variation with pulsation periods an order-of-magnitude longer \citep{green03}. The variability in these long period sdBV stars, (or V1093\,Her stars) show frequencies below about 500$\mu$Hz; this variability is caused by gravity modes ($g$-modes) as suggested by theoretical models \citep{fontaine03}. The $g$-modes are established deeper in the stars than the pressure modes. Hybrid stars, pulsating in both kind of modes were also found observationally \citep{schuh06}. As they exhibit both types of modes, models of these stars can give better constraints on their interiors since the pulsations probe both the deeper layers as well as layers nearer the surface.

\renewcommand{\thefigure}{\arabic{figure}\alph{subfigure}}
\setcounter{subfigure}{1}

\begin{figure}
\includegraphics[width=83mm]{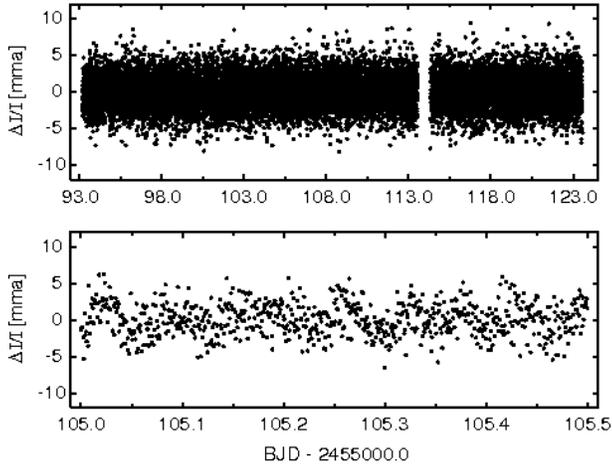}
\caption{Photometric data for KIC\,7668647. Top panel shows the entire span of the data (with outliers removed); the bottom panel zooms in on part of the data to show representative variations.}
\label{phot_7668647}
\end{figure}

Observations of long period sdBV stars, including detection of new pulsators, are not easy from the ground. Relatively long period photometric variations (with only a few cycles per night) can be affected by variable sky transparency.  To make definitive determination of the pulsation frequency requires extended photometric campaigns, preferably at several sites widely spaced in longitude to reduce day/night aliasing. However, {\em Kepler} provides long--duration continuous, homogenous, evenly spaced time-series photometry, making it an ideal instrument for asteroseismology. The {\em Kepler} science goals, mission design, and overall performance are reviewed by \cite{borucki10}, \cite{koch10} and \cite{jenkins10}.

\addtocounter{figure}{-1}
\addtocounter{subfigure}{1}

\begin{figure}
\includegraphics[width=83mm]{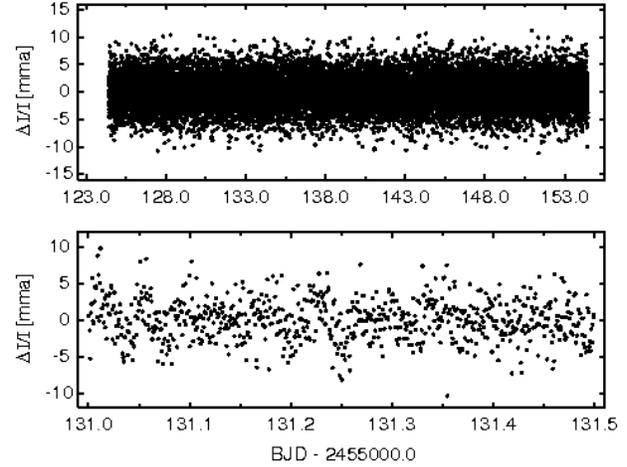}
\caption{Same as Fig. \ref{phot_7668647}, but for KIC\,10001893.}
\label{phot_10001893}
\end{figure}

Objects classified as hot subdwarfs located in field of view of the {\em Kepler} photometer were observed during the survey phase in the first year of the mission. As the spacecraft is rolled every 3 months, the survey phase was divided into 4 quarters. Analysis of the observations of compact stars from the first half of the survey phase (Q0, Q1 and Q2) is described in the following papers: \cite[][Paper I]{ostensen10b}, \cite[][Paper II]{kawaler10a}, \cite[][Paper III]{reed10}, \cite[][Paper IV]{vangrootel10}, \cite[][Paper V]{kawaler10b} and \cite{ostensen10c}.

In this paper we focus on pulsating sdB stars discovered with data from the last 6 months of the survey phase (Q3 and Q4). These are (ordered by quarters): KIC\,7668647, KIC\,8302197, KIC\,10001893, KIC\,10553698 and KIC\,11558725. In the Q3 data we found 4 sdBV of V1093\,Her class, while in the Q4 dataset we found a fifth object.  All stars presented in this paper are spectroscopically confirmed sdB stars (see Paper VI). In Table \ref{spec_par} we provide their spectroscopic parameters. They were obtained by fitting spectra to model grids in order to derive effective temperature, surface gravity and photospheric helium abundance, which was an indication whether an object is a hot subdwarf or not. Spectroscopic data along with fitting procedure are provided in the leading paper for the final half of the survey phase \citep[][Paper VI]{ostensen11}.

\addtocounter{figure}{-1}
\addtocounter{subfigure}{1}

\begin{figure}
\includegraphics[width=83mm]{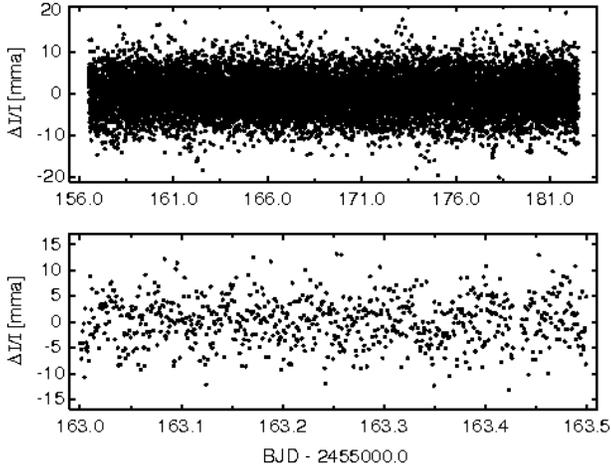}
\caption{Same as Fig. \ref{phot_7668647}, but for KIC\,8302197. Here, the variability is not as obvious. Compared to all stars in this sample, the average noise level is the largest, and amplitudes of detected peaks are the smallest.}
\label{phot_8302197}
\end{figure}

Although we did not find any predominantly $p$-mode pulsators from Q3 or Q4, we detected many peaks in the short period regime in many targets. However, these are few, compared with the large number of peaks identified in the longer period $g$-mode range. Moreover, {\em Kepler} data have a variety of instrumental issues causing some artefacts to appear in the photometric data at frequencies comparable to the $p$-modes seen in sdBV stars. Some of these artefacts are related to the long-cadence (LC) readout timings, are well-characterised, and their frequencies are known to high precision (as described in details in the {\em Kepler} Data Release Notes and Gilliland et al. 2010). We believe, though,  that some single peaks in the short period regime may also be associated with as--yet unidentified artefacts, since they are seen in more than one object.  Based on this, we treat all peaks with frequencies higher than 1000$\mu$Hz with caution. On the other hand if they are confirmed to be intrinsic to the stars, then all stars presented in this paper are hybrid pulsators.

We have also found peaks in the ``transition region'' between the $p$- and $g$-modes (i.e. between 500 and 2000 $\mu$Hz). These peaks were observed in the sdBV stars Balloon\,090100001, KIC\,2697388 by \cite{baran09} and \cite{reed10}, respectively. Their authenticity also needs confirmation.

The overall survey strategy and properties of the candidates have been described in prior papers;  Paper VI presents all selected targets for Q3 and Q4.  Further analysis of the pulsation spectra of these stars (and several from the first part of the survey) is described by \cite[][Paper VIII]{reed11}, who attempt to disentangle asymptotic behaviour of $g$-mode pulsations in V1093\,Her class stars observer by {\em Kepler} and {\em CoRoT}.

\addtocounter{figure}{-1}
\addtocounter{subfigure}{1}

\begin{figure}
\includegraphics[width=83mm]{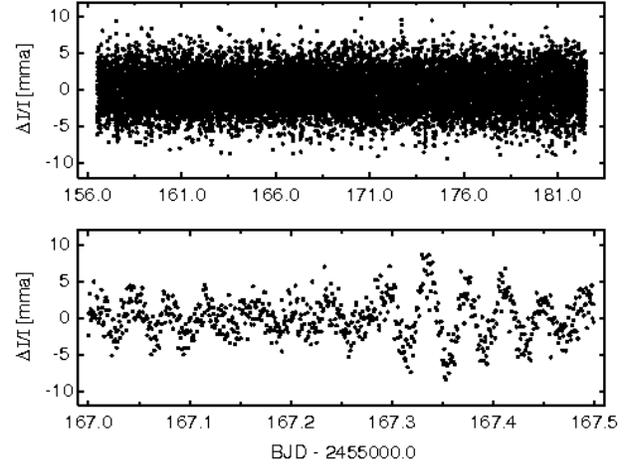}
\caption{Same as Fig. \ref{phot_7668647}, but for KIC\,11558725.}
\label{phot_11558725}
\end{figure}

\section{Observations}
Data from Q3 cover 3 months, starting from BJD\,=\,2455092.7123 (Q3.1) and ending at 2455182.0065 (Q3.3). Q4 covers the next 3 months from BJD\,=\,2455184.8679 (Q4.1) to 2455274.7137 (Q4.3). One safe-mode event occurred during Q3. As a result, data for two days (between BJD\,=\,2455154 and 2455156) are not available. As it happened at the beginning of the third month in Q3, it did not cause any gaps in the data presented here. However, the third month of Q3 (Q3.3) was truncated by those 2 days, slightly reducing the frequency resolution as well as slightly increasing the noise level in the amplitude spectrum as a result of less data (and fewer pulsation cycles). Two stars in this paper are affected by the safe-mode event: KIC\,8302197 and KIC\,11558725.

Beside this safe-mode event, the {\em Kepler} spacecraft lost its fine pointing control twice.  The first was a brief interval between BJD\,=\,245099.91 and 2455100.04, which was quickly corrected. The second, longer pointing control error fell between BJD\,=\,2455113.05 and 2455113.83, and required correction via ground command. Only data on KIC\,7668647 were affected by this fault. A log of the photometric data is presented in Table \ref{phot_log}.

\addtocounter{figure}{-1}
\addtocounter{subfigure}{1}

\begin{figure}
\includegraphics[width=83mm]{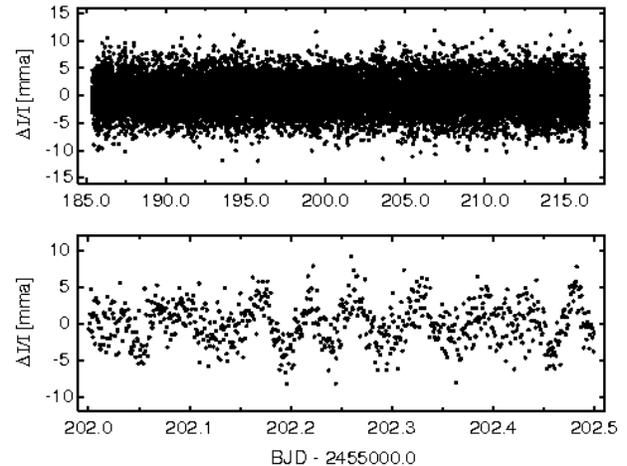}
\caption{Same as Fig. \ref{phot_7668647}, but for KIC\,10553698.}
\label{phot_10553698}
\end{figure}

All data were obtained through the {\em Kepler} Asteroseismic Science Consortium (KASC) -- see \cite{gilliland10}. The KASC data include raw as well as calibrated fluxes. Calibration takes into account all known instrumental issues as well as contamination from nearby objects. Since spectroscopic parameters of stars in the {\em Kepler} database are, generally, not precise, we believe that this contamination may not be evaluated correctly. That is why we decided to use raw fluxes for all stars in our analysis and not to account for contamination. This may lead to decreased amplitudes of detected modes by a factor of $(1-c)$, where $c$ is a contamination factor.  Future processing, supported by ground based observation of the relevant fields, will allow more accurate determination of the correction factor.
\begin{table}
 \centering
 \begin{minipage}{83mm}
  \caption{Spectroscopic parameters with their errors of five stars analyzed in this paper.}
  \label{spec_par}
  \begin{tabular}{@{}ccccc@{}}
  \hline
  KIC           &  $K_{\rm p}$ & T$_{\rm eff} [K]$ & $\log g$ & $\log$ N$_{He}$/N$_{H}$ \\
\hline
 7668647  & 15.4                & 27,700(300)       &  5.45(4) & -2.5(1)           \\
10001893 & 15.8                & 26,700(300)       &  5.30(4) & -2.9(1)           \\
 8302197  & 16.4                & 26,400(300)       &  5.32(4) & -2.7(1)           \\
11558725 & 14.9                & 27,400(200)       &  5.37(3) & -2.8(1)           \\
10553698 & 15.1                & 27,600(400)       &  5.33(5) & -2.9(2)           \\
\hline
\end{tabular}
\end{minipage}
\end{table}

We analyzed short cadence (SC) data. One SC data point is a sum of nine integration of 6.02\,s followed by 0.52\,s readout resulting in 58.85\,s time resolution. This in turn gives 8496.356\,$\mu$Hz as a Nyquist frequency. Preparing data for Fourier analysis, first we removed all SC points flagged with errant times or fluxes.  Then, all remaining points (if necessary) were subject to de-trending. This was done via a cubic spline fit calculated on 0.3\,days subsets and subtracted from original data. De-trending with this interval will suppress any frequencies below 23\,$\mu$Hz. In the next step, we removed all outliers using 4$\sigma$ clipping. Only then, the data were subject to Fourier analysis. All reduced data for each star, with shorter representative segments of the light curves, are presented in Fig.\ref{phot_7668647}-\ref{phot_10553698}.

\addtocounter{figure}{0}
\addtocounter{subfigure}{-4}

\begin{figure}
\includegraphics[width=83mm]{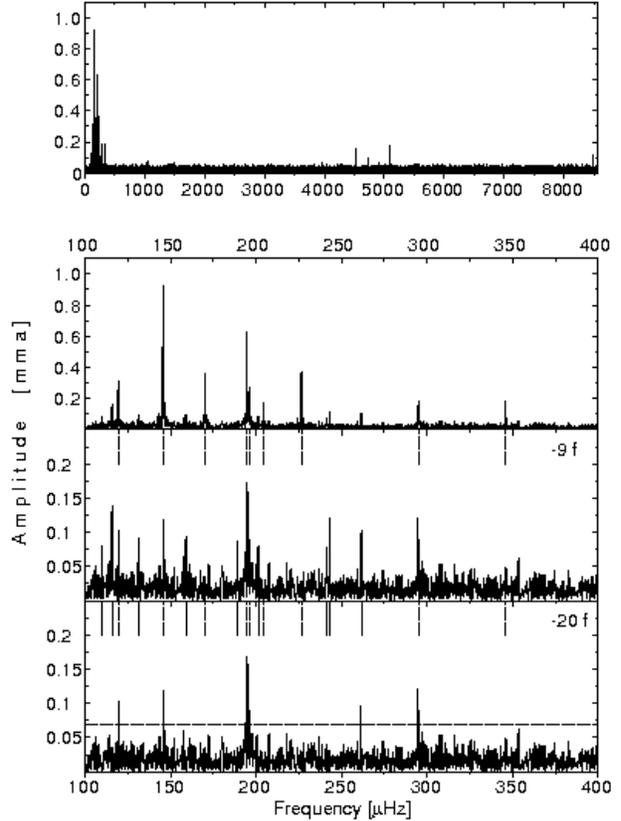}
\caption{The amplitude spectrum of KIC\,7668647. The top panel shows theoriginal amplitude spectrum up to the Nyquist frequency, while the second panel from the top shows the low frequency domain. The third panel shows the residual amplitude spectrum after removal of the 9 peaks with the highest amplitudes (those frequencies are marked with short dashed lines). The bottom panel shows the Fourier transform of the residuals after 20 peaks have been removed. Short solid lines denote the additional 11 periodicities removed from the data.  The horizontal line represents S/N\,=\,4 threshold at 0.068\,mma.}
\label{ft_7668647}
\end{figure}

\begin{table*}
 \centering
  \caption{Results of the pre-whitening process for KIC\,7668647. The phases are given at mean epoch BJD\,=\,2455108.261682. The numbers in parentheses are the errors of the last digits. In all tables presenting list of detected peaks, horizontal lines divide whole range of frequencies into three regions: $g$-mode region (defined by the bulk of peaks), "transition region" between $g$-modes and 2000\,$\mu$Hz and frequencies - above 2000\,$\mu$Hz.}
  \label{freq_7668647}
  \begin{tabular}{@{}cccccc@{}}
  \hline
ID   & Frequency [$\mu$Hz]  & Period[s]       & Amplitude [mma] & phase [rad] & S/N \\
\hline
f$_{\rm 1}$   & 109.963(39)  &   9094.0(3.2)     &   0.074(14) & 2.11(19)   &   4.3 \\
f$_{\rm 2}$   & 115.878(21)  &   8629.8(1.6)     &   0.138(14) & 4.59(10)   &   8.1 \\
f$_{\rm 3}$   & 119.514(10)  &   8367.2(0.7)     &   0.307(14) & 1.234(45)   &  18.0 \\
f$_{\rm 4}$   & 131.477(31)  &   7605.9(1.8)     &   0.093(14) & 5.54(15)   &   5.5 \\
f$_{\rm 5}$   & 145.709(3)    &   6862.99(0.14) &   0.922(14) & 4.562(15) &  54.2 \\
f$_{\rm 6}$   & 159.026(31)  &   6288.3(1.2)     &   0.092(14) & 6.01(15)   &   5.4 \\
f$_{\rm 7}$   & 170.130(8)    &   5877.86(0.28) &   0.347(14) & 5.774(40) &  20.4 \\
f$_{\rm 8}$   & 189.076(34)  &   5288.9(1.0)     &   0.086(14) & 0.33(16)   &   5.1 \\
f$_{\rm 9}$   & 194.522(5)    &   5140.81(0.13) &   0.640(14) & 1.802(22) &  37.6 \\
f$_{\rm 10}$ & 196.084(10) &   5099.86(0.26) &   0.289(14) & 2.769(48) &  17.0 \\
f$_{\rm 11}$ & 201.319(36) &   4967.2(1.0)      &   0.080(14) & 3.16(17)   &   4.7 \\
f$_{\rm 12}$ & 204.513(18) &   4889.66(0.43) &   0.163(14) & 4.70(9)   &   9.6 \\
f$_{\rm 13}$ & 226.532(8)   &   4414.39(0.16) &   0.372(14) & 4.648(37) &  21.9 \\
f$_{\rm 14}$ & 241.143(35) &   4146.9(0.6)     &   0.083(14) & 4.12(17)   &   4.9 \\
f$_{\rm 15}$ & 243.157(24) &   4112.57(0.41) &   0.123(14) & 4.70(11)   &   7.2 \\
f$_{\rm 16}$ & 261.736(28) &   3820.64(0.41) &   0.103(14) & 2.90(14)   &   6.1 \\
f$_{\rm 17}$ & 295.183(16) &   3387.73(0.18) &   0.185(14) & 2.66(8)   &  10.9 \\
f$_{\rm 18}$ & 345.795(16) &   2891.89(0.13) &   0.187(14) & 2.81(8)   &  11.0 \\
\hline
f$_{\rm 19}$ &1043.00(37) &    958.77(0.03) &   0.079(14) & 2.41(18) &   4.6 \\
\hline
f$_{\rm 20}$ &4737.812(31) &    211.068(0.001) &   0.095(14) & 2.12(15) &   5.6 \\
\hline
\end{tabular}
\end{table*}

\section{Amplitude spectra}
We used Fourier analysis to identify candidate pulsation frequencies.  We then did a nonlinear least squares fit including each periodicity in the form $A_i \sin (\omega_i t + \phi_i)$.  This was done in an iterative process, following the standard pre-whitening procedure and continued until all peaks with amplitudes satisfying signal-to-noise ratio (hereafter: S/N) $\geq$ 4 rule had been removed. Further details and examples can be found in Papers II, III, and V.

{\bf KIC\,7668647}
The amplitude spectrum of this object is dominated by a number of peaks between 100 and 400\,$\mu$Hz and is shown in Fig. \ref{ft_7668647}. The top panel presents all frequencies up to the Nyquist frequency, while the second panel isolates the frequency range between 100 and 400 $\mu$Hz. Peaks located in this frequency range result from $g$-mode pulsation. We can easily identify more than ten peaks in the original amplitude spectrum. The amplitudes less than, or on the order of, 1\,mma are typical for V1093\,Her stars.

\addtocounter{figure}{-1}
\addtocounter{subfigure}{1}

\begin{figure}
\includegraphics[width=83mm]{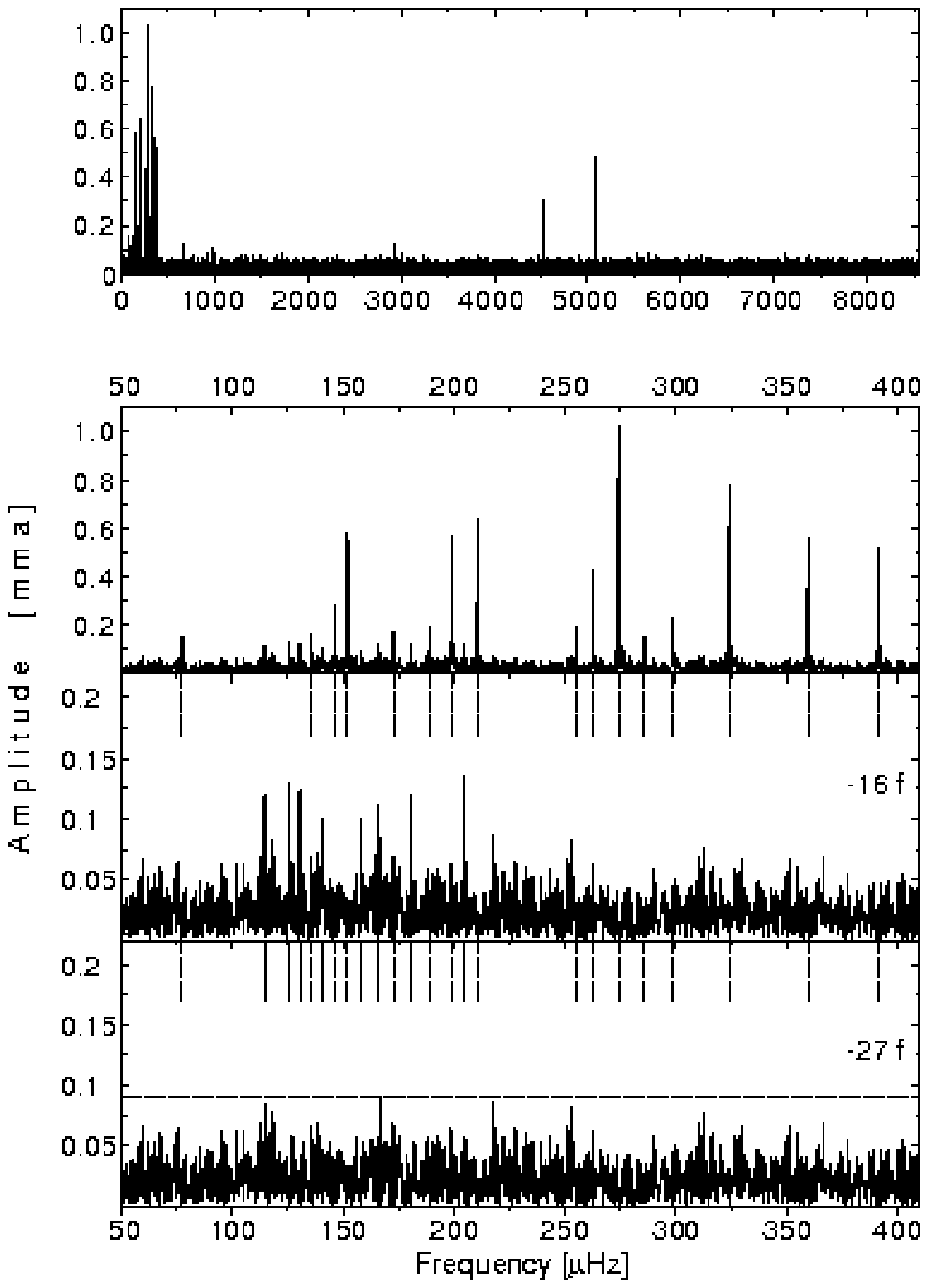}
\caption{Same as Fig. \ref{ft_7668647} but for KIC\,10001893, with the third panel showing residuals after removing 16 periodicities, and the bottom panel after 27 periodicities removed.  The \ S/N\,=\,4 threshold is at 0.088\,mma.}
\label{ft_10001893}
\end{figure}

Aside from these peaks, the top panel of Fig. \ref{ft_7668647} shows several peaks at higher frequencies.  These are signatures of the artefacts at multiples of the LC frequency of 566.4 $\mu$Hz (7$^{\rm th}$, 8$^{\rm th}$ and 14$^{\rm th}$ harmonics) and two other peaks at about 1000 and 4700\,$\mu$Hz. The peak at 1000\,$\mu$Hz is located in a "transition region" where peaks in other sdBV stars (mentioned in Section \ref{introduction}) were identified, including analysis of ground based data for Balloon\,090100001). The frequency of 4700\,$\mu$Hz for the other peak is typical for $p$-modes in sdB pulsators. We should note however that such single peaks are observed for almost all sdBV stars detected with {\em Kepler} photometry, and the ones here were barely detected in our data. Although they are not associated with any known artefacts, they should be treated with caution now.   With more data coming (runs of several months up to one year) we will be able to say with greater certainty whether or not they are intrinsic to the star. If confirmed, this object will be exposed as a hybrid pulsator. 

These two peaks were taken into account in our analysis as they are isolated and do not affect determination of the parameters of the lower-frequency $g$-modes. In total we detected 20 periodicities with S/N$\geq$4. All pre-whitened frequencies are listed in Table \ref{freq_7668647}. The bottom panel of Fig. \ref{ft_7668647} still contains peaks with significant amplitude. However, all those peaks lie close to already--removed periodicities. This may indicate amplitude/phase variation of the periodicities, or that there are a number of nearly--degenerate oscillation modes that remain unresolved in this 30-day observation. We did not attempt to pre-whiten those residual peaks. The mean noise level in the final residual amplitude spectrum is 0.017\,mma.

\begin{table*}
 \centering
  \caption{Results of the pre-whitening process for KIC\,10001893. The phases are given at mean epoch BJD\,=\,2455139.431910. The numbers in parentheses are the errors of the last digits.}
  \label{freq_10001893}
  \begin{tabular}{@{}cccccc@{}}
  \hline
ID   & Frequency[$\mu$Hz] & Period[s]        & Amplitude [mma] & phase [rad] & S/N \\
\hline
f$_{\rm 1}$   &  77.522(24)  &  12899.5(4.0)    &   0.157(18) & 4.44(12) &   7.2 \\
f$_{\rm 2}$   & 114.370(32) &   8743.6(2.5)     &   0.120(18) & 2.04(15) &   5.4 \\
f$_{\rm 3}$   & 125.910(30) &   7942.2(1.9)     &   0.128(18) & 2.82(14) &   5.8 \\
f$_{\rm 4}$   & 130.551(31) &   7659.8(1.8)     &   0.125(18) & 5.90(15) &   5.7 \\
f$_{\rm 5}$   & 135.338(24) &   7388.9(1.3)     &   0.159(18) & 3.99(11) &   7.2 \\
f$_{\rm 6}$   & 140.592(38) &   7112.8(1.9)     &   0.101(18) & 1.51(18) &   4.6 \\
f$_{\rm 7}$   & 146.087(13) &   6845.2(0.6)     &   0.297(18) & 3.71(6) &  13.5 \\
f$_{\rm 8}$   & 152.054(7)   &   6576.61(0.30) &   0.593(18) & 1.716(31) &  27.0 \\
f$_{\rm 9}$   & 158.001(39) &   6329.1(1.6)     &   0.100(18) & 5.49(18) &   4.5 \\
f$_{\rm 10}$ & 165.301(34) &   6049.6(1.2)     &   0.113(18) & 4.21(16) &   5.1 \\
f$_{\rm 11}$ & 172.628(22) &   5792.8(0.8)     &   0.179(18) & 1.94(10) &   8.1 \\
f$_{\rm 12}$ & 180.499(32) &   5540.2(1.0)     &   0.121(18) & 5.65(15) &   5.5 \\
f$_{\rm 13}$ & 189.072(20) &   5289.0(0.6)     &   0.197(18) & 1.30(9) &   8.9 \\
f$_{\rm 14}$ & 199.112(7)   &   5022.30(0.18) &   0.575(18) & 5.721(32) &  26.1 \\
f$_{\rm 15}$ & 204.713(28) &   4884.9(0.7)     &   0.136(18) & 5.76(13) &   6.2 \\
f$_{\rm 16}$ & 210.688(6)   &   4746.35(0.14) &   0.644(18) & 4.233(28) &  29.3 \\
f$_{\rm 17}$ & 255.153(20) &   3919.22(0.31) &   0.189(18) & 6.14(10) &   8.6 \\
f$_{\rm 18}$ & 262.954(9)   &   3802.95(0.13) &   0.426(18) & 4.039(42) &  19.4 \\
f$_{\rm 19}$ & 274.301(4)   &   3645.63(0.05) &   1.023(18) & 3.512(18) &  46.5 \\
f$_{\rm 20}$ & 286.025(23) &   3496.20(0.28) &   0.168(18) & 3.09(11) &   7.6 \\
f$_{\rm 21}$ & 298.656(16) &   3348.33(0.18) &   0.234(18) & 2.63(8) &  10.6 \\
f$_{\rm 22}$ & 323.988(5)   &   3086.53(0.05) &   0.782(18) & 4.452(23) &  35.6 \\
f$_{\rm 23}$ & 359.685(7)   &   2780.21(0.05) &   0.564(18) & 3.619(32) &  25.6 \\
f$_{\rm 24}$ & 391.367(7)   &   2555.15(0.05) &   0.525(18) & 4.925(34) &  23.9 \\
\hline
f$_{\rm 25}$ & 679.150(30) &   1472.43(0.07) &   0.127(18) & 1.46(14) &   5.8 \\
f$_{\rm 26}$ & 977.465(37) &   1023.05(0.04) &   0.103(18) & 1.53(18) &   4.7 \\
\hline
f$_{\rm 27}$ &2925.794(30) &    341.788(0.004) &   0.130(18) & 0.14(14) &   5.9 \\
\hline
\end{tabular}
\end{table*}

{\bf KIC\,10001893}
As in the case of the previous object, the amplitude spectrum of KIC\,10001893, presented in the top panel of Fig. \ref{ft_10001893}, is also dominated by a number of peaks in the low frequency region. Two harmonics of the LC artefact (7$^{\rm th}$ and 8$^{\rm th}$) appear in the high frequency range. Three remaining high-frequency peaks lie slightly above the detection threshold. Two of them are found in a "transition region" and one in the $p$-mode region. The latter, if real, might be a signature of hybridity. In the low frequency region we detected 24 peaks. The frequencies and amplitudes (of order of 1\,mma) suggest that these peaks are associated with $g$-modes, so it is another V1093\,Her star. The list of detected peaks is given in Table \ref{freq_10001893}.

The bottom panel of Fig. \ref{ft_10001893} displays the final residual amplitude spectrum with all 27 detected peaks removed. The horizontal line represents a detection threshold of 4 times the mean noise level, which for this star is 0.088\,mma. As is clearly seen, no peaks are left in the residuals at this level or above. This denotes that all peaks were resolved and no amplitude/phase variations on time scale of the run duration are present in this star.

\addtocounter{figure}{-1}
\addtocounter{subfigure}{1}

\begin{figure}
\includegraphics[width=83mm]{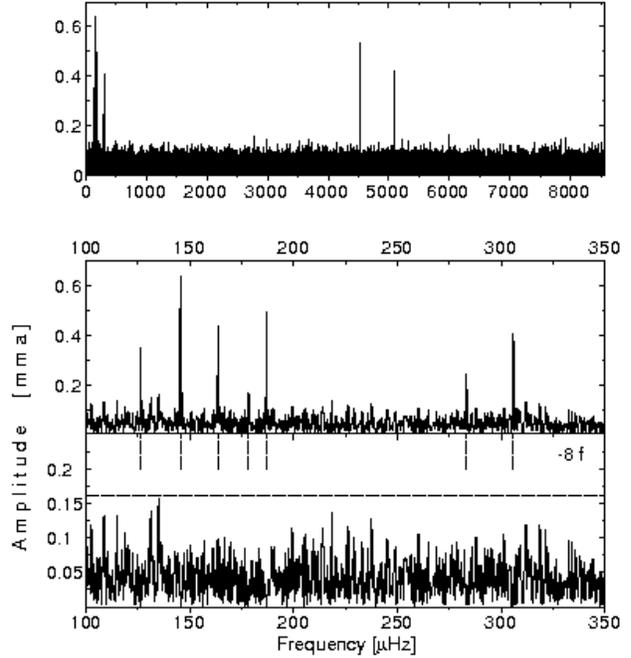}
\caption{Amplitude spectra of KIC\,8302197. The top panel shows the original amplitude spectrum up to the Nyquist frequency. The middle panel shows the relevant low-frequency domain, while the bottom panel shows the residual amplitude spectrum after removal of all 8 peaks. Short vertical dashed lines indicate the seven removed frequencies. The S/N\,=\,4 threshold at 0.164\,mma is indicated by the dashed horizontal line in the bottom panel.}
\label{ft_8302197}
\end{figure}

{\bf KIC\,8302197}
This star has the smallest number of peaks in the amplitude spectrum (Fig.\ref{ft_8302197}) in this sample. We found only 7 peaks in the low frequency region, shown in the middle panel of Fig.\ref{ft_8302197}, and one at high frequency with an amplitude above the detection threshold. All peaks are listed in Table \ref{freq_8302197}. We can easily associate these seven peaks with gravity modes and identify this object as a V1093\,Her-type star. The amplitude of the highest peaks is only 0.647\,mma.

The single peak at high frequency does not match any of the known LC artefact frequencies, so we include it in our solution. Note that it was detected at S/N\,=\,4 so its detection is marginal.  Confirmation will require more data. As in the case of the previous objects, if its authenticity is confirmed this object will be another hybrid candidate. The amplitude spectrum also contains two strong LC artifacts (7$^{\rm th}$ and 8$^{\rm th}$ harmonics).

\begin{table*}
 \centering
  \caption{Results of the pre-whitening process for KIC\,8302197. The phases are given at mean epoch BJD\,=\,2455169.508834. The numbers in parentheses are the errors of the last digits.}
  \label{freq_8302197}
  \begin{tabular}{@{}cccccc@{}}
  \hline
ID   & Frequency [$\mu$Hz]  & Period[s]     & Amplitude [mma]  & phase [rad] & S/N \\
\hline
f$_{\rm 1}$ & 126.295(23) &   7918.0(1.4)      &   0.354(33) & 5.3(0.8)      &   8.6 \\
f$_{\rm 2}$ & 145.588(12) &   6868.7(0.6)      &   0.647(33) & 2.94(0.42) &  15.8 \\
f$_{\rm 3}$ & 163.683(18) &   6109.4(0.7)      &   0.443(33) & 5.5(0.6)      &  10.8 \\
f$_{\rm 4}$ & 178.335(46) &   5607.4(1.5)      &   0.172(33) & 2.2(1.6)      &   4.2 \\
f$_{\rm 5}$ & 186.989(16) &   5347.91(0.46) &   0.497(33) & 1.0(0.5)    &  12.1 \\
f$_{\rm 6}$ & 283.243(33) &   3530.54(0.41) &   0.241(33) & 4.7(1.1)      &   5.9 \\
f$_{\rm 7}$ & 305.834(20) &   3269.75(0.21) &   0.409(33) & 0.1(0.7)      &  10.0 \\
\hline
f$_{\rm 8}$ &6002.352(48) &    166.601(0.001) &   0.165(33) & 4.8(1.6) &   4.0 \\
\hline
\end{tabular}
\end{table*}

The bottom panel of Fig.\ref{ft_8302197} shows the final residual amplitude spectrum with all 8 detected peaks removed.  This process did not reveal unresolved peaks with significant amplitudes. The time span of the data is sufficient to resolve all modes down to the detection threshold, which for this star is equal to 0.164\,mma. 

\addtocounter{figure}{-1}
\addtocounter{subfigure}{1}

\begin{figure*}
\includegraphics[width=110mm]{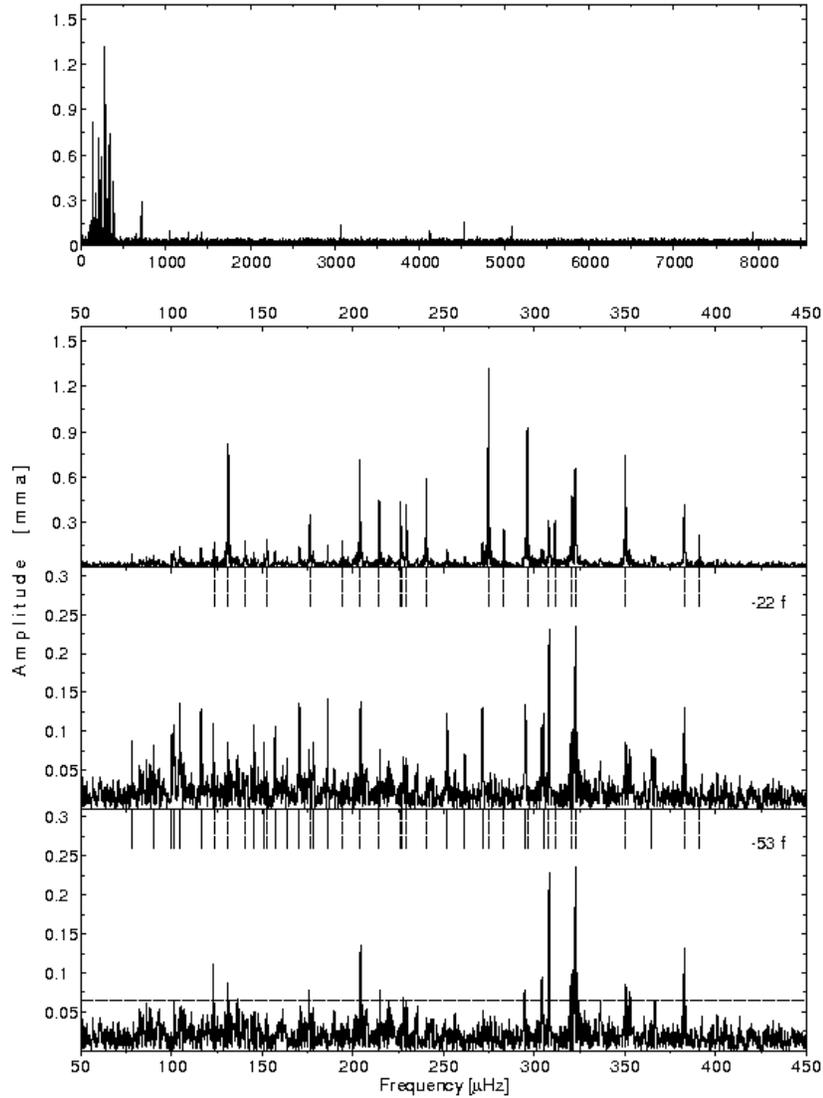}
\caption{Same as Fig. \ref{ft_7668647} but for KIC\,11558725.  The third panel is the Fourier transform after 22 peaks are removed, and the bottom panel after 53 peaks removed. Short solid vertical lines mark the other 31 removed peaks. The S/N\,=\,4 threshold is at 0.064\,mma.}
\label{ft_11558725}
\end{figure*}

{\bf KIC\,11558725}
The amplitude spectrum of this object, shown in Fig.\ref{ft_11558725}, contains the largest number of detected peaks, listed in Table \ref{freq_11558725}. A total of 41 peaks are revealed in the low frequency region between 50 and 400\,$\mu$Hz.  There are also quite a few in a "transition region" (9 peaks) as well as at high frequencies (3 peaks). Only two of them are found at high confidence. The S/N ratio for f$_{\rm 43}$ and f$_{\rm 44}$ is 12 and 17.4, respectively. It indicates that they are very significant and may be intrinsic to the star. We have to keep in mind, however, that there is still chance that these peaks might be caused by a contaminating star; the  contamination factor for KIC\,11558725 is 0.269 as currently listed in the {\em Kepler} Input Catalog.  Unlike other objects presented in this paper, KIC\,11558725 shows a significant number of peaks at higher frequencies. It is very likely that some (if not all) of them are real $p$-modes. This make the object best candidate for hybrid star.

After removal of all 53 periodicities, the amplitude spectrum still contains residual signals, as can be seen in the bottom panel of Fig.\ref{ft_11558725}. In particular, there are two strong peaks between 300 and 350\,$\mu$Hz. These two, along with other outstanding peaks above the horizontal dashed line at the detection threshold of 0.064\,mma, were not removed, as they lie at frequencies within the frequency resolution limit of stronger peaks already removed from data. Either they are intrinsic to the star and caused by amplitude/phase changes, or result from beating between unresolved periodicities. With the longer data sets we expect to obtain, we hope to resolve this issue.  We also found three harmonics of LC artifact (6$^{\rm th}$, 8$^{\rm th}$ and 13$^{\rm th}$).

\begin{table*}
 \centering
  \caption{Results of the pre-whitening process for KIC\,11558725. The phases are given at mean epoch BJD\,=\,2455169.508302. The numbers in parentheses are the errors of the last digits.}
  \label{freq_11558725}
  \begin{tabular}{@{}cccccc@{}}
  \hline
ID   & Frequency [$\mu$Hz]         & Period[s]         & Amplitude [mma]    & phase [rad] & S/N \\
\hline
f$_{\rm 1}$   &  78.132(37)   &  12798.9(6.1)     &   0.087(13) & 1.1(6)      &   5.5 \\
f$_{\rm 2}$   &  90.131(39)   &  11095.0(4.8)     &   0.082(13) & 4.5(7)      &   5.1 \\
f$_{\rm 3}$   & 100.091(36)  &   9990.9(3.6)      &   0.094(13) & 0.5(6)      &   5.9 \\
f$_{\rm 4}$   & 101.002(33)  &   9900.8(3.2)      &   0.103(13) & 4.6(6)      &   6.5 \\
f$_{\rm 5}$   & 104.524(24)  &   9567.2(2.2)      &   0.132(13) & 5.49(41) &   8.3 \\
f$_{\rm 6}$   & 116.106(24)  &   8612.8(1.8)      &   0.131(13) & 0.31(42) &   8.2 \\
f$_{\rm 7}$   & 123.435(17)  &   8101.4(1.1)      &   0.187(13) & 4.04(29) &  11.7 \\
f$_{\rm 8}$   & 131.148(4)    &   7624.97(0.23) &   0.832(13) & 2.84(7)    &  52.0 \\
f$_{\rm 9}$   & 140.418(18)  &   7121.6(0.9)      &   0.180(13) & 3.19(30) &  11.3 \\
f$_{\rm 10}$ & 145.076(30)  &   6892.9(1.4)      &   0.108(13) & 5.48(50) &   6.8 \\
f$_{\rm 11}$ & 150.918(36)  &   6626.1(1.6)      &   0.090(13) & 3.7(6)     &   5.6 \\
f$_{\rm 12}$ & 152.612(19)  &   6552.6(0.8)      &   0.173(13) & 4.35(32)&  10.8 \\
f$_{\rm 13}$ & 156.957(30)  &   6371.2(1.2)      &   0.107(13) & 5.4(5)     &   6.7 \\
f$_{\rm 14}$ & 163.770(46)  &   6106.1(1.7)      &   0.069(13) & 6.2(8)     &   4.3 \\
f$_{\rm 15}$ & 170.378(24)  &   5869.3(0.8)      &   0.135(13) & 1.58(41) &   8.4 \\
f$_{\rm 16}$ & 176.204(10)  &   5675.24(0.32) &   0.342(13) & 3.47(16) &  21.4 \\
f$_{\rm 17}$ & 178.229(37)  &   5610.8(1.2)      &   0.087(13) & 1.9(6)     &   5.4 \\
f$_{\rm 18}$ & 186.044(22)  &   5375.1(0.7)      &   0.142(13) & 3.96(38) &   8.9 \\
f$_{\rm 19}$ & 194.148(20)  &   5150.7(0.5)      &   0.161(13) & 0.99(34) &  10.1 \\
f$_{\rm 20}$ & 203.747(5)    &   4908.05(0.12) &   0.709(13) & 2.17(8)    &  44.3 \\
f$_{\rm 21}$ & 214.413(7)    &   4663.90(0.15) &   0.435(13) & 2.08(13) &  27.2 \\
f$_{\rm 22}$ & 226.192(8)    &   4421.02(0.16) &   0.416(14) & 3.82(14) &  26.0 \\
f$_{\rm 23}$ & 226.996(20)  &   4405.36(0.39) &   0.167(14) & 5.88(34) &  10.4 \\
f$_{\rm 24}$ & 229.400(8)    &   4359.20(0.15) &   0.425(13) & 2.29(13) &  26.5 \\
f$_{\rm 25}$ & 240.364(5)    &   4160.36(0.09) &   0.593(13) & 5.30(9)   &  37.0 \\
f$_{\rm 26}$ & 251.902(26)  &   3969.80(0.41) &   0.123(13) & 3.46(44) &   7.7 \\
f$_{\rm 27}$ & 261.621(45)  &   3822.32(0.66) &   0.071(13) & 0.4(8)     &   4.4 \\
f$_{\rm 28}$ & 271.288(25)  &   3686.12(0.34) &   0.129(13) & 0.97(42) &   8.1 \\
f$_{\rm 29}$ & 274.686(2)    &   3640.52(0.03) &   1.321(13) & 5.450(41)   &  82.5 \\
f$_{\rm 30}$ & 283.222(13)  &   3530.80(0.16) &   0.248(13) & 0.49(22) &  15.5 \\
f$_{\rm 31}$ & 295.079(24)  &   3388.92(0.28) &   0.137(13) & 1.92(40) &   8.6 \\
f$_{\rm 32}$ & 296.096(4)    &   3377.28(0.05) &   0.930(13) & 2.66(6)   &  58.2 \\
f$_{\rm 33}$ & 305.200(26)  &   3276.54(0.28) &   0.124(13) & 0.05(44) &   7.8 \\
f$_{\rm 34}$ & 307.604(10)  &   3250.93(0.11) &   0.314(13) & 1.34(18) &  19.6 \\
f$_{\rm 35}$ & 311.336(10)  &   3211.96(0.10) &   0.331(13) & 5.77(17) &  20.7 \\
f$_{\rm 36}$ & 320.842(7)    &   3116.80(0.07) &   0.473(13) & 4.82(12) &  29.6 \\
f$_{\rm 37}$ & 322.493(5)    &   3100.84(0.05) &   0.667(13) & 4.00(8)   &  41.7 \\
f$_{\rm 38}$ & 350.187(4)    &   2855.62(0.03) &   0.749(13) & 5.14(7)   &  46.8 \\
f$_{\rm 39}$ & 364.732(42)  &   2741.74(0.32) &   0.076(13) & 2.0(7)     &   4.8 \\
f$_{\rm 40}$ & 382.609(8)    &   2613.63(0.05) &   0.427(13) & 5.29(13) &  26.7 \\
f$_{\rm 41}$ & 390.899(16)  &   2558.21(0.10) &   0.206(13) & 6.16(26) &  12.9 \\
\hline
f$_{\rm 42}$ & 658.183(38)  &   1519.33(0.09) &   0.084(13) & 4.2(7)      &   5.2 \\
f$_{\rm 43}$ & 710.810(17)  &   1406.85(0.03) &   0.192(13) & 1.47(28) &  12.0 \\
f$_{\rm 44}$ & 712.361(12)  &   1403.78(0.02) &   0.279(13) & 5.26(20) &  17.4 \\
f$_{\rm 45}$ &1039.500(50) &    962.00(0.05)  &   0.068(13) & 4.6(9)      &   4.3 \\
f$_{\rm 46}$ &1040.622(46) &    960.96(0.04)  &   0.077(13) & 1.3(8)      &   4.8 \\
f$_{\rm 47}$ &1041.727(38) &    959.94(0.04)  &   0.089(13) & 1.1(6)      &   5.5 \\
f$_{\rm 48}$ &1270.344(37) &    787.19(0.02)  &   0.086(13) & 3.1(6)      &   5.4 \\
f$_{\rm 49}$ &1370.516(44) &    729.65(0.02)  &   0.072(13) & 4.2(8)      &   4.5 \\
f$_{\rm 50}$ &1423.021(35) &    702.73(0.02)  &   0.091(13) & 4.4(6)      &   5.7 \\
\hline
f$_{\rm 51}$ &3073.449(23) &    325.367(0.002) &   0.137(13) & 0.45(40) &   8.6 \\
f$_{\rm 52}$ &3075.084(31) &    325.194(0.003) &   0.104(13) & 3.3(5)      &   6.5 \\
f$_{\rm 53}$ &4114.660(32) &    243.033(0.002) &   0.100(13) & 3.2(5)      &   6.2 \\
\hline
\end{tabular}
\end{table*}

\addtocounter{figure}{-1}
\addtocounter{subfigure}{1}

\begin{figure*}
\includegraphics[width=110mm]{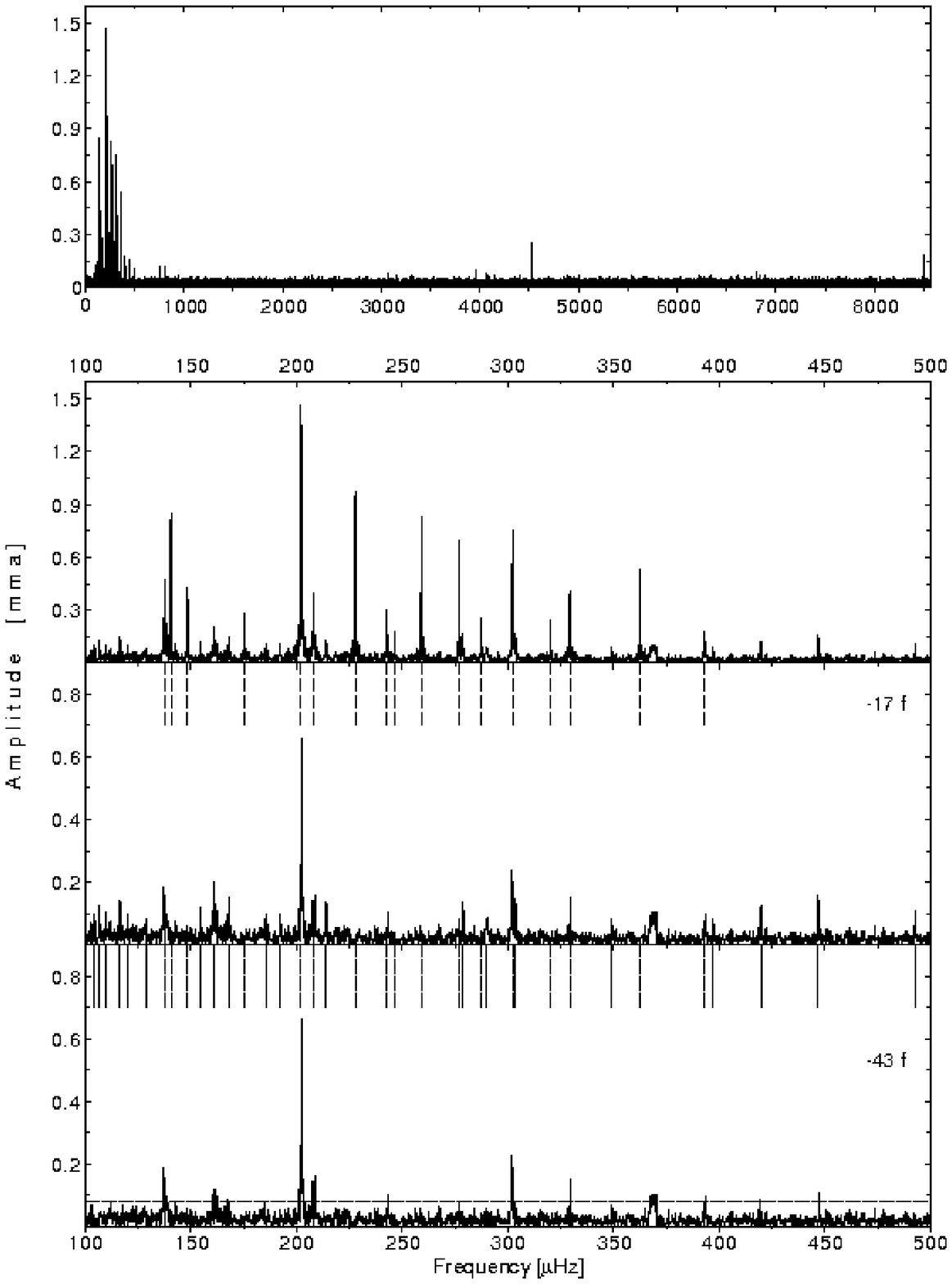}
\caption{Same as Fig. \ref{ft_7668647} but for KIC\,10553698. 17 peaks with the highest amplitudes were removed prior to producing the Fourier transform shown in the third panel, and 43 peaks were removed to produce the spectrum in the fourth panel.  The S/N\,=\,4 threshold is at 0.076\,mma.}
\label{ft_10553698}
\end{figure*}

{\bf KIC\,10553698}
This object also shows a very rich amplitude spectrum. As seen in Fig.\ref{ft_10553698}, the spectrum is dominated by forest of peaks in the low frequency region characteristic of V1093\,Her pulsators. We can easily identify at least 17 peaks with the highest amplitudes.  Another 26 peaks (43 in total) with lower amplitudes were detected above the detection threshold of 0.076\,mma and removed from data.  We present our results in Table \ref{freq_10553698}.

Even after all these peaks have been pre-whitened, there is still residual signal present in the final residual amplitude spectrum (bottom panel in Fig.\ref{ft_10553698}). The highest residual peak occurred very close to the dominant mode frequency.  To sort out this problem we likely need more data to either resolve close frequencies or monitor amplitude/phase changes.

\renewcommand{\thefigure}{\arabic{figure}}

\begin{figure}
\includegraphics[width=83mm]{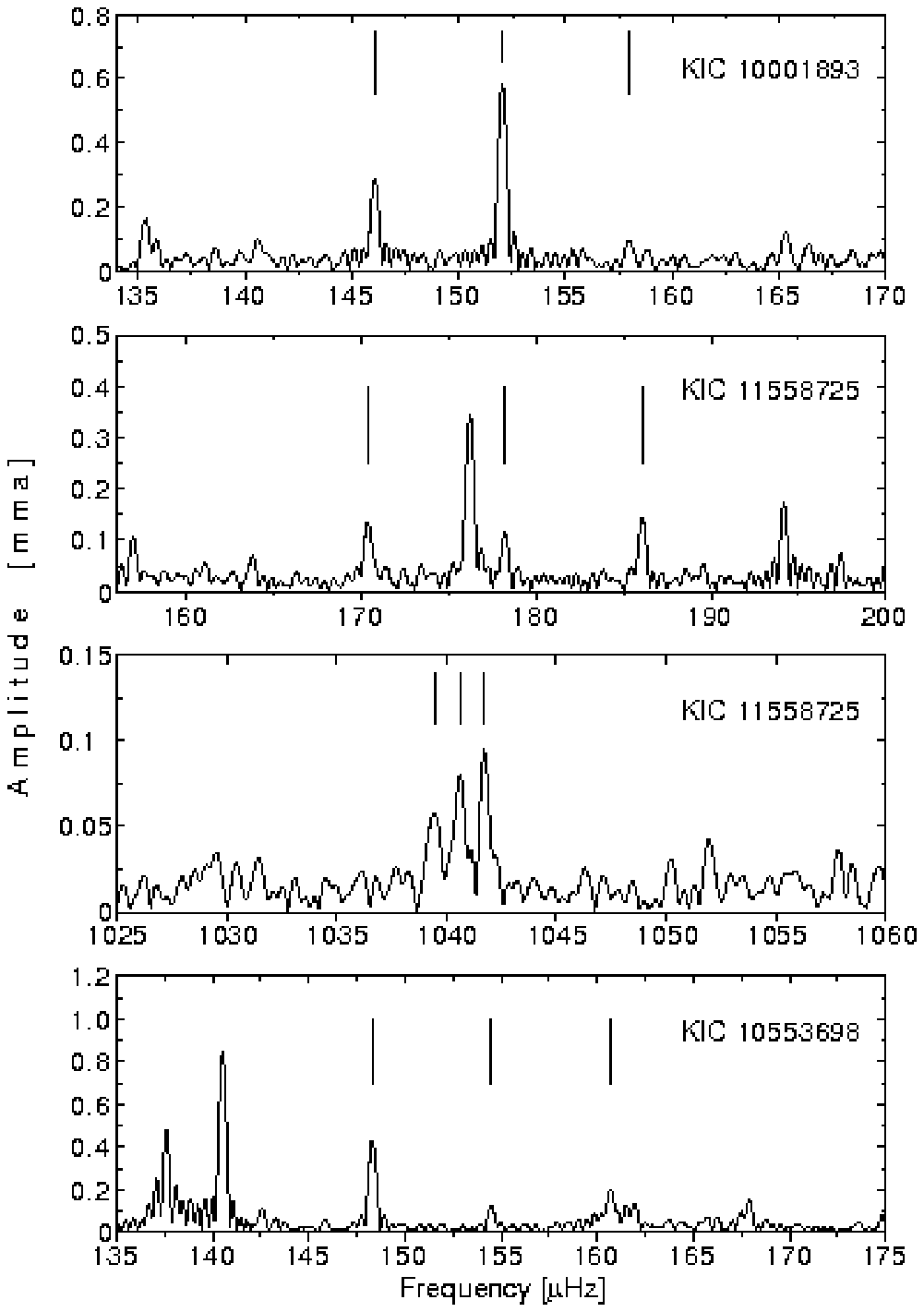}
\caption{Best candidates for triplets in three stars. Vertical solid lines denote the location of peaks which form the triplets.}
\label{triplets}
\end{figure}

Other than $g$-modes, we found 4 signatures of LC artifacts (6$^{\rm th}$, 7$^{\rm th}$, 10$^{\rm th}$ and 14$^{\rm th}$ harmonics) and six unidentified peaks. Four peaks are located in the ``transition region'' and two are at high frequencies. All of them were detected with low significance and their authenticity has yet to be confirmed.

In the bottom panel of Fig.\ref{ft_10553698} we can also notice a broad set of peaks at 360\,$\mu$Hz. Although their amplitudes are above the detection threshold and are not residual peaks of any formerly removed, we did not attempt to pre-whiten this region. This complex pattern is seen in many other compact stars observed with {\em Kepler}. It was seen in Q0 through Q4 data, and most frequently in the Q3 data release. Although the cause of this feature is unknown and it is seen in a few stars, we assume that these peaks are caused by some instrumental effects and were not included in our analysis.

Comparing results for the two stars KIC\,11558725 and KIC\,10553698 we found that each star has a peak at nearly the same frequency: f$_{\rm 51}$ in the former star and f$_{\rm 42}$ in the latter differ by 0.4$\mu$Hz, which is comparable to the formal resolution of each run. We believe that these peaks are not intrinsic to the stars (either target or contaminating ones) and likely to be caused by instrumental issues. Presumably f$_{\rm 52}$ in KIC\,11558725 is a part of the complex structure and is also caused by the same instrumental effect. Since we cannot be absolutely sure about their origin yet we keep them in our solution.  They do not influence peaks in the $g$-mode region.

\begin{table*}
 \centering
  \caption{Results of the pre-whitening process for KIC\,10553698. The phases are given at mean epoch BJD\,=\,2455200.902734. The numbers in parentheses are the errors of the last digits.}
  \label{freq_10553698}
  \begin{tabular}{@{}cccccc@{}}
  \hline
ID   & Frequency [$\mu$Hz]  & Period[s] & Amplitude [mma]    & phase [rad] & S/N \\
\hline
f$_{\rm 1}$   & 104.291(35) &   9588.6(3.2)      &   0.094(16) & 3.76(17) &   4.9 \\
f$_{\rm 2}$   & 106.562(27) &   9384.2(2.4)      &   0.123(16) & 3.29(13) &   6.5 \\
f$_{\rm 3}$   & 109.701(35) &   9115.7(2.9)      &   0.095(16) & 3.60(17) &   5.0 \\
f$_{\rm 4}$   & 116.395(23) &   8591.4(1.7)      &   0.145(16) & 1.15(11) &   7.7 \\
f$_{\rm 5}$   & 120.037(32) &   8330.8(2.2)      &   0.104(16) & 1.79(15) &   5.5 \\
f$_{\rm 6}$   & 128.775(39) &   7765.5(2.4)      &   0.084(16) & 1.00(19) &   4.4 \\
f$_{\rm 7}$   & 137.569(7)   &   7269.08(0.37) &   0.475(16) & 3.869(34) &  25.0 \\
f$_{\rm 8}$   & 140.513(4)   &   7116.78(0.20) &   0.842(16) & 0.973(19) &  44.3 \\
f$_{\rm 9}$   & 148.313(8)   &   6742.50(0.36) &   0.434(16) & 1.964(37) &  22.8 \\
f$_{\rm 10}$ & 154.486(27) &   6473.1(1.1)      &   0.121(16) & 4.91(13) &   6.4 \\
f$_{\rm 11}$ & 160.672(16) &   6223.9(0.6)      &   0.201(16) & 4.08(8) &  10.6 \\
f$_{\rm 12}$ & 167.904(21) &   5955.8(0.7)      &   0.154(16) & 6.19(10) &   8.1 \\
f$_{\rm 13}$ & 175.342(12) &   5703.14(0.39) &   0.285(16) & 5.72(6) &  15.0 \\
f$_{\rm 14}$ & 185.728(34) &   5384.2(1.0)      &   0.097(16) & 3.40(17) &   5.1 \\
f$_{\rm 15}$ & 192.202(34) &   5202.9(0.9)      &   0.098(16) & 0.44(16) &   5.1 \\
f$_{\rm 16}$ & 201.963(2)   &   4951.40(0.05) &   1.473(16) & 0.780(11) &  77.5 \\
f$_{\rm 17}$ & 208.060(8)   &   4806.31(0.18) &   0.401(16) & 5.729(40) &  21.1 \\
f$_{\rm 18}$ & 213.971(24) &   4673.5(0.5)      &   0.139(16) & 4.05(16) &   7.3 \\
f$_{\rm 19}$ & 227.685(3)   &   4392.03(0.06) &   0.979(16) & 1.508(16) &  51.5 \\
f$_{\rm 20}$ & 242.534(11) &   4123.13(0.19) &   0.308(16) & 4.07(5) &  16.2 \\
f$_{\rm 21}$ & 246.393(19) &   4058.56(0.31) &   0.173(16) & 5.46(9) &   9.1 \\
f$_{\rm 22}$ & 259.086(4)   &   3859.72(0.06) &   0.828(16) & 0.679(19) &  43.6 \\
f$_{\rm 23}$ & 276.896(5)   &   3611.46(0.07) &   0.696(16) & 0.133(23) &  36.7 \\
f$_{\rm 24}$ & 278.732(24) &   3587.68(0.31) &   0.138(16) & 0.47(12) &   7.3 \\
f$_{\rm 25}$ & 287.148(14) &   3482.52(0.17) &   0.245(16) & 3.42(7) &  12.9 \\
f$_{\rm 26}$ & 290.031(39) &   3447.91(0.46) &   0.084(16) & 0.45(19) &   4.4 \\
f$_{\rm 27}$ & 302.226(4)   &   3308.78(0.04) &   0.766(16) & 3.16(2) &  40.3 \\
f$_{\rm 28}$ & 303.565(23) &   3294.19(0.25) &   0.145(16) & 1.55(11) &   7.7 \\
f$_{\rm 29}$ & 320.011(13) &   3124.89(0.13) &   0.244(16) & 3.67(7) &  12.9 \\
f$_{\rm 30}$ & 329.286(8)   &   3036.87(0.07) &   0.411(16) & 1.242(39) &  21.6 \\
f$_{\rm 31}$ & 348.780(39) &   2867.14(0.32) &   0.083(16) & 4.05(19) &   4.4 \\
f$_{\rm 32}$ & 362.430(6)   &   2759.15(0.05) &   0.537(16) & 4.817(30) &  28.2 \\
f$_{\rm 33}$ & 392.894(18) &   2545.22(0.12) &   0.179(16) & 1.15(9) &   9.4 \\
f$_{\rm 34}$ & 397.115(39) &   2518.16(0.25) &   0.084(16) & 2.96(19) &   4.4 \\
f$_{\rm 35}$ & 419.673(27) &   2382.81(0.15) &   0.124(16) & 3.03(13) &   6.5 \\
f$_{\rm 36}$ & 446.767(21) &   2238.30(0.11) &   0.157(16) & 5.24(10) &   8.3 \\
f$_{\rm 37}$ & 492.869(30) &   2028.94(0.12) &   0.109(16) & 0.12(15) &   5.7 \\
\hline
f$_{\rm 38}$ & 755.331(34) &   1323.92(0.06) &   0.099(16) & 5.77(16) &   5.2 \\
f$_{\rm 39}$ & 756.996(28) &   1321.01(0.05) &   0.119(16) & 6.06(14) &   6.3 \\
f$_{\rm 40}$ & 807.468(29) &   1238.44(0.04) &   0.118(16) & 4.17(14) &   6.2 \\
f$_{\rm 41}$ & 808.590(31) &   1236.72(0.05) &   0.110(16) & 1.54(15) &   5.8 \\
\hline
f$_{\rm 42}$ &3073.803(40) &    325.330(0.004) &   0.083(16) & 2.19(19) &   4.4 \\
f$_{\rm 43}$ &4069.917(40) &    245.705(0.002) &   0.082(16) & 0.71(20) &   4.3 \\
\hline
\end{tabular}
\end{table*}

\section{Discussion}

With the available sets of frequencies for each star, we can now examine them for the the asymptotic signature of $g$-mode pulsations, the coherence of the observed pulsations, and the asteroseismic effects of rotation and binarity.  In this section, we identify the possible signature of rotation (equally split multiplets), and discuss binarity (low-frequency photometric variation) and the possible hybrid behaviour of these stars.  We also demonstrate the asymptotic signature of $g$-mode pulsation, equal period spacings for consecutive overtones of the same degree $l$, but leave more detailed discussion of this property to 
a companion paper (Paper VIII).

\subsection{Rotationally split multiplets}

For nonradial pulsation in spherically stars, all modes with nonzero values of the azimuthal quantum number $m$ have the same frequency as the $m$\,=\,0 mode. Azimuthal symmetry can be destroyed, lifting the $m$ degeneracy, by processes such as rotation or global magnetic fields. Since all stars rotate, we expect to see multiplets in the amplitude spectra, with the splitting within multiplets reflecting the rotation frequency. If all components of a multiplet are manifest, it will have 2$l$+1 components. If the rotation is sufficiently slow (with a rotation period that is much longer than the oscillation period) the central frequency ($m$\,=\,0) will not be affected by rotation while all other components with nonzero $m$ values will be shifted by $\nu_{n,l,m}=\nu_{n,l,0}+m\Omega(1-C_{n,l})$ \citep{ledoux51} in the simplest case of solid--body rotation at a frequency $\Omega$.  The Ledoux parameter $C_{n,l}$ is an integral quantity that depends on the radial dependence of the oscillation eigenfunctions through the star.  Identification of multiplets in the amplitude spectrum can help to derive the rotation period, and is very useful for mode identification. 

Thus far we detected multiplets in only a few V361\,Hya class stars \citep[e.g.][]{reed04,baran09}. The very rich amplitude spectrum of V1093\,Her class stars detected with {\em Kepler} hold the potential to expose multiplets in these stars.  Without any obvious evidence of these stars being in close binaries (with synchronized rotation), we do not have any prior expectation of what splittings to look for (unlike the stars discussed in Paper V).  Therefore, we constrained our search to rotation frequencies that range from the minimum that can be resolved with 30 days of data (on the slow rotation side) to rotation periods that would not produce rotational broadening of spectral lines greater than the limits placed via spectroscopy. We assume that Ledoux parameter for $g$-modes (C$_{n,1}$\,=\,0.5, and C$_{n,2}$\,=\,0.17 -- see Paper V).  These constraints limit us to splittings ranging from 0.825 to 13.5\,$\mu$Hz, which correspond to rotation periods between 7\,days and 10.3\,hrs, respectively. We looked for triplets only, and we assumed that their symmetries can deviate within the formal frequency uncertainty, which is about 0.02\,$\mu$Hz.

We do not find any potential candidates for triplets in KIC\,7668647 or KIC\,8302197. There are few possibilities in the other three stars. KIC\,10001893 shows two triplet candidates. They are both slightly asymmetrical. One consists of f$_{\rm 7}$, f$_{\rm 8}$ and f$_{\rm 9}$ with splittings equal to 5.97 and 5.95\,$\mu$Hz, while the other triplet is formed by f$_{\rm 9}$, f$_{\rm 10}$ and f$_{\rm 11}$ with splittings equal to 7.30 and 7.33\,$\mu$Hz. The f$_{\rm 9}$ peak appears in both triplets, which is clearly inconsistent with a simple rotational splitting interpretation. Since the amplitude of the modes in the 5.96\,$\mu$Hz-spaced triplet are significantly larger, we have more confidence that it is a true multiplet.

In KIC\,11558725 we found two other candidates. These are: f$_{\rm 15}$, f$_{\rm 17}$ and f$_{\rm 18}$, having splittings of 7.85 and 7.82\,$\mu$Hz, and f$_{\rm 45}$, f$_{\rm 46}$ and f$_{\rm 47}$ with splittings equal to 1.12 and 1.11\,$\mu$Hz.  These clearly are not consistent with a uniform rotation period within the star, and even models with differential rotation (with depth) rarely show this extreme difference \citep{kaw05}. For KIC\,10553698 we find a candidate triplet formed by f$_{\rm 9}$, f$_{\rm 10}$ and f$_{\rm 11}$, with splittings of 6.17 and 6.19\,$\mu$Hz. The amplitude spectra surrounding these triplets are presented in Fig. \ref{triplets}.

\renewcommand{\thefigure}{\arabic{figure}\alph{subfigure}}
\setcounter{subfigure}{1}

\begin{figure*}
\includegraphics[width=150mm]{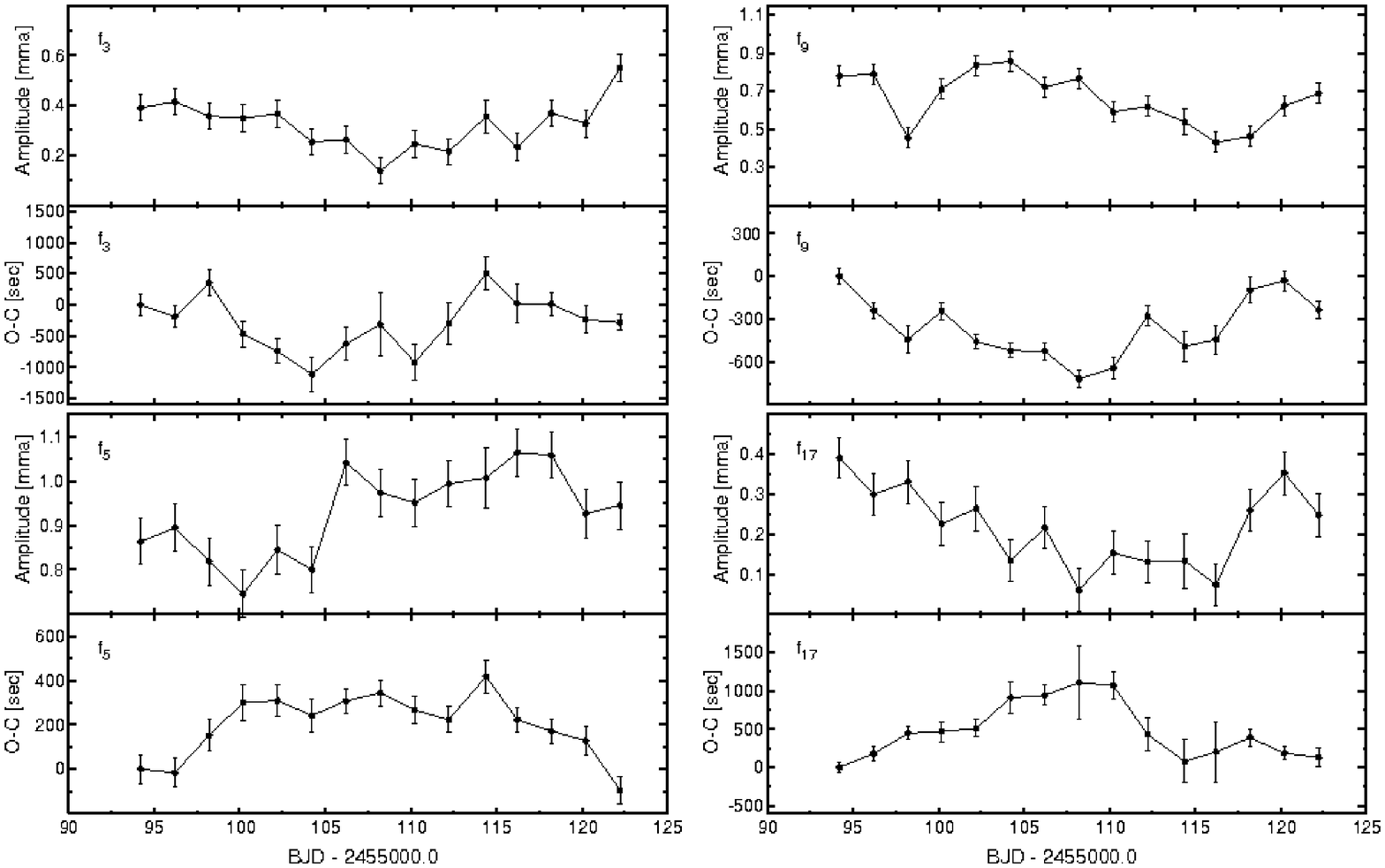}
\caption{Amplitude and phase variation in KIC\,7668647. We used a least-squares fit to the frequencies indicated in each panel and fit 2-day subsets of the data to obtain amplitudes and phase. Each pair of plots shows the amplitude as a function of time, and (O-C), expressed in seconds, to represent the observed phase compared to the expected value. }
\label{ampl_ph_7668647}
\end{figure*}

Although we found a few candidates for rotationally split triplets, it is hard to explain why only very few modes within a given star are rotationally split. But there are similarities with other sdBV stars. For example, \cite{baran09} found only two multiplets among more than fifty periodicities in Balloon\,090100001, and those involved the largest-amplitude modes. Assuming we are seeing real triplets here, a rough estimate of the average rotation period for those three stars is about 20\,hrs.

\subsection{Binarity}

\cite{maxted01} and \cite{morales06}, among others, argue that approximately half of all sdB stars are members of binary systems. A signature of binarity can be found in photometric observations (via eclipses, modulation from reflection off of a low--mass companion, and/or through ellipsoidal variations) or through radial velocity variations measured via spectroscopy.  Many binary sdB stars detected in photometry, e.g. HW \,Vir \citep{wood93}, NY\,Vir \citep{kilkenny98} having beautiful eclipses features or KPD\,1930+2752 \citep{billeres00} showing ellipsoidal variation. Recently another binary sdB stars have been found using {\em Kepler} data \citep{kawaler10b, ostensen10c}. The first describes two pulsating sdB stars that are reflection-effect binaries; the second is a pulsator in an eclipsing binary. 

Analysis of the five stars presented here does not reveal any evidence for either the reflection effect or ellipsoidal variation or eclipses.  We note that the de-trending described in Sect. 2 might mask some of that evidence, so we also analysed the data prior to de-trending. While detailed radial velocity time series have not been performed, there is no evidence for composite spectra in these targets.  Therefore, we have no evidence of binarity for these stars.   We note that with a longer time baseline, timing variations of the principal pulsation modes can be used to reveal a companion via reflex orbital motion, as was done by \citet{silv07} to reveal a planetary companion to V391\,Peg.

\subsection{Do these stars show $p$-mode pulsations?}

As mentioned in Sect. 1, the two classes of pulsating sdB stars overlap in the H--R diagram.  
On the hotter end, we find $p$-mode pulsators (the V361\,Hya class). Typically, these are hotter than $T_{\rm eff}$\,$\approx$\,30\,kK.  On the cool side we have $g$-mode pulsators (V1093\,Her) with typical $T_{\rm eff}$ of about 25\,kK. 
In the overlap region, stars can show both $g$-mode and $p$-mode pulsations.  Since these modes sample different regions of the stars, they are particularly potent targets for asteroseismic investigation. Prior to the {\em Kepler} discoveries, these stars were relatively rare, with only 5 examples found in ground-based studies.  In all of those cases, the stronger pulsations are in the $p$-mode frequency range. 
The lone {\em Kepler} $p$-mode pulsator had a single low-amplitude oscillation at low frequencies (Paper II), suggesting that it behaves like the already-known hybrid pulsators.

For the {\em Kepler} sdBV stars from the first part of the survey, we found that many of the $g$-mode pulsators showed suggestive evidence for shorter period pulsations in the $p-$mode regime (Paper III) at lower amplitude.  This pattern is not seen in any of the sdBV pulsators observed using ground-based data.  This is likely a signal-to-noise issue.  The $g$-modes are low amplitude to begin with, and at problematic frequencies for adequate ground-based observing; the possible $p$-modes we see are at even lower amplitude.  

\addtocounter{figure}{-1}
\addtocounter{subfigure}{1}

\begin{figure*}
\includegraphics[width=150mm]{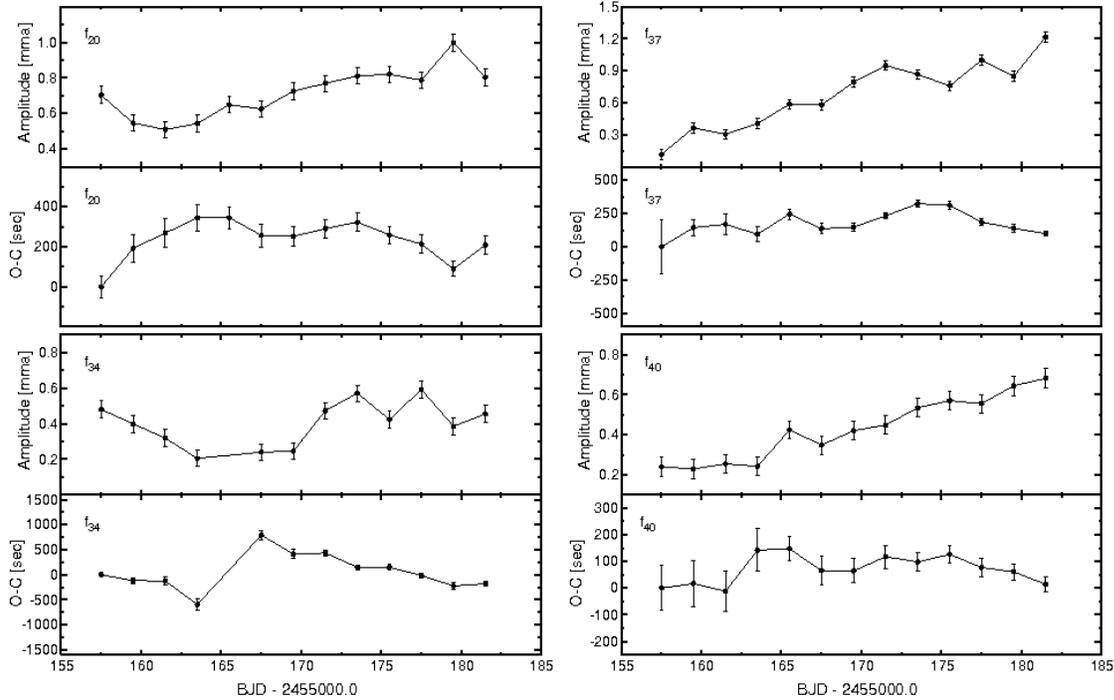}
\caption{Same as Fig. \ref{ampl_ph_7668647} but for KIC\,11558725.}
\label{ampl_ph_11558725}
\end{figure*}

Based on their effective temperatures,  almost all the potential {\em Kepler} hybrid candidates \citep[five stars in this paper together with two stars from paper by][]{reed10} have a typical T$_{\rm eff}$ close to 27\,kK. The only exception is the slightly-cooler KIC\,5807616 with only one barely detected peak at high frequencies. Their location in $\log g$ and $T_{\rm eff}$ diagram is right between hybrid stars dominated by $p$-modes and V1093\,Her class stars. This would be consistent with the idea that there is a transition temperature between $p$-mode pulsations at the hot end and $g$-modes at the cool end; the $p$-mode dominant stars are in general hotter than the {\it Kepler} $g$-mode dominant hybrids. If the single high-frequency periodicities are confirmed, this could explain their presence.

The issue of isolated $p$-modes is addressed in more detail in one of these stars in the modeling work by \cite{charp11} of the star KIC\,2697388. Their theoretical models suggest that one peak at high frequency is likely intrinsic to the star, despite its low amplitude. Longer coverage should provide a definitive test of the reality of this mode.

\subsection{Amplitude and phase stability}

While pre-whitening we assumed that each periodicity maintained a fixed frequency, amplitude and phase. This does not have to be the case, however, since the pulsation can experience amplitude and phase modulation through nonlinear interactions, or growth or decay of mode energy over the course of the run.  Changes in amplitude and/or phase can produce a broadening of the peaks in the amplitude spectrum, or produce multiple closely-spaced peaks.  Manifestation as multiple peaks could mimic the appearance of multiple (fully coherent) pulsation modes, so resolving the amplitude spectrum requires care to distinguish between these possibilities.

For some stars, pre-whitening leaves behind no signals above our noise limits in the residual amplitude spectra.  When no residual signal at the frequency of a removed peak remains, we can be sure that the peak was resolved and the frequency, amplitude, and phase are constant over the course of data run. Looking back at the residual spectra obtained in our analysis we can see that no residual signals remain after pre-whitening the data from KIC\,8302197 and KIC\,10001893. Removal of the tabulated peaks leaves no significant signal behind.

However, for three stars, residual peaks remain after pre-whitening the dominant periodicities.  Their residual spectra contain additional signal at the frequencies of formerly removed peaks. These may be closely spaced coherent modes, or the effects of amplitude and phase modulation. For KIC\,7668647 this extra signal remains in four places with relatively small amplitudes.  For two other stars, KIC\,1158725 and KIC\,10553698, significant peaks do remain after prewhitening.  In the former, there are two strong peaks slightly above 300\,$\mu$Hz while there is one strong residual peak in the latter. In both cases there are other peaks with smaller significance but still above the detection threshold.

There is a correlation between the amplitude and phase variations that is characteristic of beating between closely-spaced coherent oscillations (i.e. the phase jumps by $180^{\circ}$ during the amplitude minimum in a beat between two equal--amplitude periodicities).  If the amplitude and/or phase vary irregularly and with no specific correlations, then that might suggest that the modes are experiencing stochastic changes.  Since we have continuous data that span about 30\,days we can use shorter subsets of the data to look for amplitude and/or phase variations, as was done in Paper II.  

\addtocounter{figure}{-1}
\addtocounter{subfigure}{1}

\begin{figure*}
\includegraphics[width=150mm]{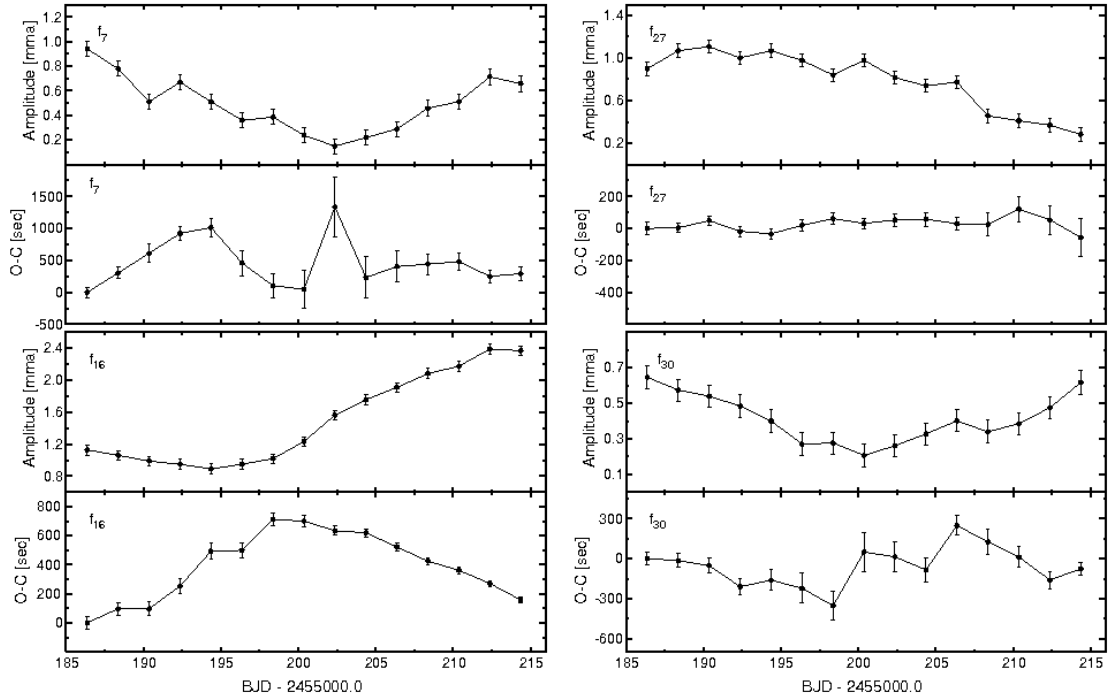}
\caption{Same as Fig.\ref{ampl_ph_7668647} but for KIC\,10553698.}
\label{ampl_ph_10553698}
\end{figure*}

To examine the phase and amplitude stability at a given frequency, we first prewhitened by all of the other significant periodicities found for that star. The prewhitened data were then divided into shorter, equal-length, contiguous runs. For each of these shorter runs, we fixed the frequency at the best value for the mode in question, and fitted the amplitude and phase. Choosing the length of subsets requires balance between time resolution and amplitude and phase uncertainties.  We performed this analysis with different subsets, and found that 2\,day samples provided the best compromise. We used non-linear least-squares to fit one sine wave to each individual subset keeping frequency fixed for each peak to determine the pulsation amplitude and phase (using a fixed but arbitrarily zero point). We repeated this fitting only for four highest residual peaks for each of those three stars. The result of this analysis is presented in Fig.\ref{ampl_ph_7668647}-\ref{ampl_ph_10553698}. 

Fig. \ref{ampl_ph_7668647} shows that the amplitude of this star changes at a significant level for all four tested frequencies.  In all of these frequencies, there is no clear correlation between the amplitude and phase modulations.  For example, there are no abrupt phase jumps in any of the modes tested; though phase variations occur, they are all much less than one cycle, and they are not correlated with a minimum in pulsation amplitude.  Thus the variations appear to be caused by random variation in the parameters of this mode. The variations seen in three of the four modes shine in Fig. \ref{ampl_ph_11558725} appear to be stochastic as well, though in $f_{34}$ there appears to be a jump in phase around day 167, when that mode is at minimum amplitude.  The period of this mode is 3250 seconds, and the phase jump is about 1600 seconds  -- suggesting that there is beating present near this mode.

KIC\,10553698 shows much slower, smoother variations in amplitude and phase in the four modes tested as compared to the other two stars.  In most cases, there is no correlation between the phase and amplitude variations for a given mode (i.e. no large phase changes near amplitude minima).  But as can be seen in Fig. \ref{ampl_ph_10553698}, $f_7$ shows an unusual phase variation near minimum amplitude around day 200-205.  The period of this mode is in excess of 7000 seconds, and the phase increases in that interval by 1500 seconds and then returns to the slower trend.  It is conceivable that the phase jumped by half a cycle (3500 seconds) and that the (O-C) curve should lie up near 3500 seconds in this plot if beating is present.  But our best guess is that the phase did not experience that large a change. It is notable that the slow variation in the amplitude of $f_{16}$ is reflected in the large residual peak near 200\,$\mu$Hz in the bottom panel of Fig.\ref{ft_10553698}.  Clearly, a longer observing run on this star should aid in determining the true nature of these periodicities.

\subsection{Asymptotic $g$-mode period spacings}

Theoretical consideration of the V1093\,Her class \citep{fontaine03} showed that the pulsation modes detected in these stars correspond to high order $g$-modes. As such, we could invoke asymptotic limit for $n\gg l$ in which consecutive overtones should be evenly spaced in period. This behavior has been observed in pulsating white dwarfs \citep{kawaler94}. Because of the low--amplitude signal that is characteristic of these stars, and the logistical challenges of observing these frequencies from a single ground--based observing site, this asymptotic behaviour had not been seen in V1093\,Her stars until space--based observations became available.  The first set of {\it Kepler} V1093\,Her stars were found to show clear evidence of equal period spacings, with values of approximately 145 and 250\,s reported in Paper III and in a later work (Paper VIII).

The five new detected V1093\,Her class also show evidence of evenly--spaced periods. For all stars we found average spacing in 250--270\,s range. According to the Paper III, all of the peaks matching this spacing are likely to be $l$\,=\,1 modes. There is also evidence of a second series of peaks with spacings in the  140--150\,s range, consistent with them being $l$\,=\,2 modes. This simple relation between consecutive overtones provides a determination of the degree $l$ of the modes, which will be crucial for interior modeling efforts.  Detailed analysis of period spacing on five stars presented here and stars already published in previous papers of our series is a subject of Paper VIII.

\section{Summary}

We found five new long-period pulsating sdB (V1093\,Her) stars using the {\it Kepler} photometer. The oscillations are mostly in the 100--400\,$\mu$Hz region, although we also found some single peaks at higher frequencies. The nature of those higher-frequency peaks, which are at low amplitude, remain uncertain, and we cannot rule out contamination from variable contaminating sources, or unknown spacecraft artefacts. On the other hand, assuming that peaks at high frequencies are indeed $p$-modes, the stars match the previously known T$_{\rm eff}$ - pulsation type correlation. 
We believe that more data should confirm or disprove our hypothesis and definitively answer the question about the origin of the peaks. The amplitudes of detected peaks in the low frequency region are about 1\% or less which is in similar to other known V1093\,Her stars. 

As we have not found any signatures of binarity we claim that these are single stars.  The pulsation spectra show only limited evidence of rotational splitting, but these occur with low-amplitude periodicities and therefore require continued observation to improve the signal-to-noise and look for other similar spacings in the same stars.  While many of the observed oscillations are coherent, several modes show the signature of possible stochastic amplitude and phase modulation.

This paper completes our initial analysis of the survey phase data on compact stars by {\it Kepler}. In total we found 12 new members of the V1093\,Her class (5 in this paper, 2 in Paper V, and 5 in Paper III), one V361\,Her star (Paper II) and one hybrid star \citep{ostensen10c}. Among the former two groups, many showing additional peaks in the opposite side of the amplitude spectrum. 

Our results demonstrate the potential of the {\it Kepler} spacecraft for asteroseismology  of sdB stars. Continuous coverage and the unprecedented quality of the data allowed us to detect, at high S/N, stellar pulsations in stars which would not be easily studied from the ground. The rich amplitude spectra of V1093\,Her class stars enables us to use the relation for period spacings to assign most of the detected peaks with $n$ (relative) and $l$ values (see Paper VIII). With those determinations, along with the frequencies themselves, tight constraints can be placed on models of these stars, and therefore will expand our understanding of the formation, structure, and evolution of these highly evolved stars.  We anticipate that with much longer runs of {\it Kepler} data we can detect even more peaks with lower amplitudes making period spacing tool even more efficient.

\section*{Acknowledgments}

AB gratefully appreciates funding from Polish Ministry of Science and Higher Education under project No. 554/MOB/2009/0.  MDR was partially funded by a Missouri State University Faculty Research
Grant. ACQ is supported by the Missouri Space Grant Consortium, funded by NASA. Funding for the {\em Kepler} Mission is provided by NASA's Science Mission Directorate. The authors gratefully acknowledge the {\em Kepler} Science Team and all those who have contributed to making the {\em Kepler} Mission possible.

\label{lastpage}


\begin{thebibliography}{99}
\bibitem[\protect\citeauthoryear{Baran et al.}{2009}]{baran09} Baran A., et al., 2009, MNRAS, 385, 255

\bibitem[\protect\citeauthoryear{Billeres et al.}{2000}]{billeres00} Bill\'eres M., Fontaine G., Brassard P., Charpinet S., Liebert J., Saffer R. A., 2003, ApJ, 597, 518

\bibitem[\protect\citeauthoryear{Borucki et al.}{2010}]{borucki10} Borucki W.J. et al., 2010, Science, 327, 977

\bibitem[\protect\citeauthoryear{Charpinet et al.}{1996}]{charp96} Charpinet S., Fontaine G., Brassard P., Dorman B., 1996, ApJ, 471, L103

\bibitem[\protect\citeauthoryear{Charpinet et al.}{2011}]{charp11} Charpinet S. et al., 2011, A\&A, submitted


\bibitem[\protect\citeauthoryear{D'Cruz et al.}{1996}]{dcruz96} D'Cruz N.L., Dorman B., Rood R.T., O'Connell R.W., 1996, ApJ, 466, 359

\bibitem[\protect\citeauthoryear{Fontaine et al.}{2003}]{fontaine03} Fontaine G., Brassard P., Charpinet S., Green E.M., Chayer P., Bill\`eres M., Randall S.K., 2003, ApJ, 597, 518

\bibitem[\protect\citeauthoryear{Gilliland et al.}{2010}]{gilliland10}Gilliland R.~L., Jenkins J.~M., Borucki W.~J. et al., 2010, ApJ, 713, L160

\bibitem[\protect\citeauthoryear{Green et al.}{2003}]{green03} Green E. M. et al., 2003, ApJ, 583, L31

\bibitem[\protect\citeauthoryear{Heber}{2009}]{heber09} Heber U., 2009, ARA\&A, 47, 211

\bibitem[\protect\citeauthoryear{Han et al.}{2002}]{han02} Han Z., Podsiadlowski Ph., Maxted P.F.L., Marsch T.R., Ivanova N., 2002, MNRAS, 336, 449

\bibitem[\protect\citeauthoryear{Han et al.}{2003}]{han03} Han Z., Podsiadlowski Ph., Maxted P.F.L., Marsh T.R., 2003, MNRAS, 341, 669

\bibitem[\protect\citeauthoryear{Jenkins et al.}{2010}]{jenkins10} Jenkins J. et al., 2010, ApJ, 713, L87

\bibitem[\protect\citeauthoryear{Kawaler \& Bradley}{1994}]{kawaler94} Kawaler S.D., Bradley P.A., 1994, ApJ, 427, 415

\bibitem[\protect\citeauthoryear{Kawaler \& Hostler}{2005}]{kaw05} Kawaler S.~D., Hostler S.~R., 2005, ApJ, 621, 432

\bibitem[\protect\citeauthoryear{Kawaler et al.}{2010a}]{kawaler10a} Kawaler S.D., Reed M.~D., Quint A.~C., et al., 2010a, MNRAS, 409, 1487 (Paper II)

\bibitem[\protect\citeauthoryear{Kawaler et al.}{2010b}]{kawaler10b} Kawaler S.D., Reed M.~D.,
{\O}stensen R. H., et al., 2010b, MNRAS, 409, 1509 (Paper V)

\bibitem[\protect\citeauthoryear{Kilkenny et al.}{1997}]{kilkenny97} Kilkenny D., Koen C., O'Donoghue D., Stobie R. S., 1997, MNRAS, 285, 640

\bibitem[\protect\citeauthoryear{Kilkenny et al.}{1998}]{kilkenny98} Kilkenny D., O'Donoghue D., Koen C., Lynas Gray A.E., van Wyk F., 1998, MNRAS, 296, 329


\bibitem[\protect\citeauthoryear{Koch et al.}{2010}]{koch10} Koch D.G., et al., 2010, ApJ, 713, 79


\bibitem[\protect\citeauthoryear{Ledoux}{1951}]{ledoux51} Ledoux P., 1951, ApJ, 114, 373

\bibitem[\protect\citeauthoryear{Maxted et al.}{2001}]{maxted01} Maxted P.F.L., Heber U., Marsh T.R., North R.C., 2001, MNRAS, 326, 1391

\bibitem[\protect\citeauthoryear{Morales-Rueda et al.}{2006}]{morales06} Morales-Rueda L., Maxted P.F.L., Marsh T.R., Kilkenny D., O'Donoghue D., 2006, Balt.A., 15, 187

\bibitem[\protect\citeauthoryear{{\O}stensen et al.}{2010b}]{ostensen10a} {\O}stensen R.H. et al., 2010a, A\&A, 513, 6

\bibitem[\protect\citeauthoryear{{\O}stensen et al.}{2010a}]{ostensen10b} {\O}stensen R.H., Silvotti R., Charpinet S. et al., 2010b, MNRAS, 409, 1470 (Paper I)

\bibitem[\protect\citeauthoryear{{\O}stensen et al.}{2010c}]{ostensen10c} {\O}stensen R.H. et al., 2010c, MNRAS, 408, L51


\bibitem[\protect\citeauthoryear{{\O}stensen et al.}{2011}]{ostensen11} {\O}stensen R.H. et.al., 2011, MNRAS, submitted (Paper VI)


\bibitem[\protect\citeauthoryear{Reed et al.}{2004}]{reed04} Reed M.D. et al., 2004, MNRAS, 348, 1164

\bibitem[\protect\citeauthoryear{Reed et al.}{2010}]{reed10} Reed M.D. et al., 2010, MNRAS, 409, 1496 (Paper III)

\bibitem[\protect\citeauthoryear{Reed et al.}{2011}]{reed11} Reed M.D. et al., 2011, MNRAS, submitted (Paper VIII)

\bibitem[\protect\citeauthoryear{Schuh et al.}{2006}]{schuh06} Schuh S., Heber U., Dreizler S., Heber U., O'Toole, S.J., Green E.M., Fontaine G., 2006, A\&A, 445, 31

\bibitem[\protect\citeauthoryear{Silvotti et al.}{2007}]{silv07} Silvotti R. et al., 2007, Nature 449, 189

\bibitem[\protect\citeauthoryear{Van Grootel et al.}{2010}]{vangrootel10} Van Grootel V. et al., 2010, ApJ, 718, L97 (Paper IV)

\bibitem[\protect\citeauthoryear{Wood, Zhang \& Robinson}{1993}]{wood93} Wood J.H., Zhang E.H., Robinson E.L., 1993, MNRAS, 261, 103
\end{thebibliography}
\end{document}